\documentclass[usegraphicx,usenatbib,natbib,useAMS]{mn2e}

\usepackage{times}     
\usepackage{aas_macros}
\usepackage{amssymb}   
\usepackage{multirow}

\def\halpha{\mbox{H$\alpha$}}

\def\lya{\mbox{Ly$\alpha$}}

\def\ecs{\mbox{\,erg~cm$^{-2}$~s$^{-1}$}}

\def\lesssim{\mathrel{\hbox{\rlap{\hbox{\lower4pt\hbox{$\sim$}}}\hbox{$<$}}}}
\def\gtrsim{\mathrel{\hbox{\rlap{\hbox{\lower4pt\hbox{$\sim$}}}\hbox{$>$}}}}

\DeclareRobustCommand{\ion}[2]{%
\relax\ifmmode
 \ifx\testbx\f@series
  {\mathbf{#1\,\mathsc{#2}}}\else
  {\mathrm{#1\,\mathsc{#2}}}\fi
 \else\textup{#1\,{\mdseries\textsc{#2}}}%
\fi}

\title[The M-Z relation in DLA galaxies]{Verifying the
  mass-metallicity relation in damped Lyman-$\alpha$ selected galaxies
  at $0.1<z<3.2$}

\author[L. Christensen et al.]
       {L. Christensen\thanks{lise@dark-cosmology.dk}$^1$, 
        P. M{\o}ller$^2$, J.~P.~U. Fynbo$^1$, T. Zafar$^2$\\
        $^1$ Dark Cosmology Centre, Niels Bohr Institute, University
        of Copenhagen, Juliane Maries Vej 30, 2100 Copenhagen, Denmark\\
        $^2$ European Southern Observatory, Karl-Schwarzschildstrasse
        2, D-85748 Garching bei M\"unchen, Germany
}
\date{Accepted 2014 August 21. Received 2014 August 14; in original form 2014 April 25}
      
\pubyear{2014}

\begin{document}

\maketitle

\label{firstpage}

\begin{abstract}

A scaling relation has recently been suggested to combine the galaxy
mass-metallicity (MZ) relation with metallicities of damped
Lyman-$\alpha$ systems (DLAs) in quasar spectra. Based on this
relation the stellar masses of the absorbing galaxies can be
predicted. We test this prediction by measuring the stellar masses of
12 galaxies in confirmed DLA absorber - galaxy pairs in the redshift
range 0.1~$<z<$~3.2.  We find an excellent agreement between the
predicted and measured stellar masses over three orders of magnitude,
and we determine the average offset $\langle C_{\mathrm{[M/H]}}
\rangle = 0.44 \pm 0.10$ between absorption and emission
metallicities. We further test if $C_{\mathrm{[M/H]}}$ could depend on
the impact parameter and find a correlation at the $5.5\sigma$ level.
The impact parameter dependence of the metallicity corresponds to an
average metallicity difference of $-0.022\pm0.004$ dex~kpc$^{-1}$.  By
including this  metallicity vs. impact parameter correlation in
the prescription instead of $C_{\mathrm{[M/H]}}$, the scatter reduces
to $0.39$ dex in log $M_*$. We provide a prescription how to calculate
the stellar mass ($M_*^{\mathrm{DLA}}$) of the galaxy when both the
DLA metallicity and DLA galaxy impact parameter is known.  We
demonstrate that DLA galaxies follow the MZ relation for
luminosity-selected galaxies at $z=0.7$ and $z=2.2$ when we include a
correction  for the correlation between impact parameter and
  metallicity.
\end{abstract}



\section{Introduction}
During the last decade, large galaxy surveys have provided a wealth of
information on the average properties of galaxies. Scaling relations
of various physical properties are useful for our understanding of
what is the driving parameter behind galaxy evolution. A
well-established scaling relation between the galaxy stellar mass and
the gas phase metallicity was cemented by investigations of 53,000
local galaxies in the Sloan Digital Sky Survey
\citep[SDSS;][]{tremonti04} suggesting that a primary parameter that
drives galaxy evolution is the stellar mass.  The simple
interpretation for the mass-dependence is a metal loss via galactic
winds, where galaxies with shallow potential well have more efficient
outflows. Deeper targeted surveys have extended the mass-metallicity (MZ)
relation to redshifts of $\sim$1 \citep{savaglio05}, $\sim$2
\citep{erb06}, and $\sim$3 \citep{maiolino08}, and show that for a
given stellar mass, galaxies have lower metallicities at increasing
redshifts. Other scaling relations that also evolve with redshift
involve galaxy stellar masses and star formation rates
\citep{noeske07}. The three observables (star-formation rate,
metallicity and stellar mass), have been shown to form a fundamental
relation that does not evolve with redshift \citep{lara-lopez10},
although \citet{mannucci10} hint that the relation might change at
$z>2.5$.  Further analysis on the redshift evolution requires detailed
spectroscopic data for individual high-redshift galaxies.

While the luminosity selected galaxies are well studied, long
integration times on near-IR spectrographs are necessary to measure
metallicities based on rest-frame optical emission lines from
high-redshift galaxies. Because of observational limitations,
preferentially the most massive and luminous galaxies are targeted and
thereby also relatively high-metallicity galaxies are
investigated. Metallicities below 10\% solar are rarely inferred for
galaxies at $z \gtrsim 2$ because of this selection effect. By
observing intrinsically fainter, low-mass galaxies, such as
gravitationally lensed galaxies, detailed analyses of galaxies at
$z>2$ with metallicities about 5--10\% solar are feasible
\citep{yuan09,wuyts12,christensen12,belli13}.

From quasar absorption line studies on the other hand, we know that
galaxies with much lower metallicities exist at $z \gtrsim 2$. In
particular, damped Lyman-$\alpha$ systems (DLAs), that have neutral
hydrogen column densities of log $N$(\ion{H}{i})~$>20.3$ cm$^{-2}$,
reveal metallicities that are typically 1--10\% solar
\citep[e.g.][]{pettini94,ledoux02,prochaska03}. At these redshifts,
even cases of extremely low DLA metallicities between 0.1--1\% solar
are known \citep{cooke11}.

DLAs have a corresponding relation analogous to the MZ relation for
luminosity selected galaxies. \citet{ledoux06} show that the velocity
widths of DLA metal lines and their metallicities are correlated. As
the velocity-metallicity relation has the same slope as the galaxy MZ
relation, it is appealing to use the width of the absorption lines as
a proxy for the stellar mass of the galaxies. This scaling relation
demonstrates that low-metallicity DLAs trace low-mass galaxies
that have lower luminosities than galaxies targeted in spectroscopic
surveys. The velocity-metallicity relation of DLAs evolves with
redshift \citep{ledoux06,moller13,neeleman13} again demonstrating a
lower metallicity for a given velocity (or equivalently mass) at
progressively higher redshifts.

The hypothesis that  low-metallicity DLAs trace galaxies at the
low-mass end of the mass distribution is supported by models that
include the galaxy luminosity, extension of a gas disc, metallicities
and gradients \citep{fynbo08} and numerical simulations are able to
reproduce the low DLA metallicities \citep{pontzen08}. Models suggest
that metal-rich DLAs can be found at larger projected distances
between the quasi-stellar object (QSO) and the absorbing galaxy, the
so-called impact parameter, and this hypothesis is supported by
observations \citep{krogager12}. In the remainder of this paper, we
will call the galaxy which is responsible for the damped absorption in
the QSO spectrum the `DLA galaxy'.

The origin of DLA clouds relative to the DLA galaxies has been widely
debated. Studies of the \ion{C}{ii}* fine structure line showed that
some DLAs might arise in-situ in star-forming galaxies \citep{wolfe04}
while others do not appear to be heated by very nearby star-forming
regions \citep{wolfe03}. Observations have suggested that DLA clouds
can be expelled through galactic winds \citep{noterdaeme12}, while
another proposition is that the low-metallicity of DLAs is indicative
of aggregates of pristine gas rather than gas processed by stars
\citep{bouche13,fukugita14}.  In this paper, we will not make any
assumptions regarding the structure of the stellar- and gas
distribution within DLA galaxies. The morphology of low-redshift DLA
galaxies comprise a wide selection of types: irregular,
bulge-dominated, low-surface brightness galaxies while others are
genuine galaxy spiral discs \citep{chen03}. In this paper, we simply
refer to DLA clouds as belonging to halos or extended outskirts of DLA
galaxies irrespectively of their origin.

Whereas the success rate of identifying DLA galaxies at low redshift
($z<1$) is close to 50\% \citep{chen03,rao03,chen05,rao11}, only few
spectroscopically confirmed DLA galaxies at $z>2$ have been reported. 
As the galaxies are expected to have faint continuum emission, several
searches have aimed at discovering the \lya\ emission line from the
DLA galaxies.  However, there are more failed attempts to discover the
DLA galaxies as the Lyman-$\alpha$ emission line is resonantly
scattered and easily absorbed by dust
\citep[e.g.][]{kulkarni06,christensen07}. Instead, searches for the
\halpha\ emission line would not be as hampered by dust absorption,
but observations have only revealed a few detections at $z\sim1$ and
$z\sim2$, while more upper detection limits are reported
\citep{peroux12,bouche12}.  Even though the majority of DLA galaxies
are low-mass and hence low-luminosity galaxies, \citet{moller04}
suggested that targeting the more metal-rich DLAs would imply a higher
chance of discovering the DLA galaxy. Indeed, this is confirmed by
recent observations
\citep{fynbo10,fynbo11,noterdaeme12,krogager13,fynbo13}.

Since we now have a growing sample of DLA galaxies, the next step is
to understand the relations between DLAs and the galaxies detected in
emission. Such a correlation is predicted by simulations of DLA
galaxies \citep{pontzen08}.  Combining the observed
velocity-metallicity relation of DLAs with the MZ
scaling relation from luminosity selected galaxies \citep{maiolino08},
\citet[][hereafter M13]{moller13} derive an  equation to compute
stellar masses, $M_*$, of DLA galaxies:
\begin{equation}
\log(M_*/M_{\odot}) = 1.76(\mathrm{[M/H]} + C_{\mathrm{[M/H]}} + 0.35z + 5.04),
\label{eq:pmfit}
\end{equation}
where [M/H] is the measured DLA absorption metallicity and $z$ is the
redshift. The remaining coefficient, $C_{\mathrm{[M/H]}}$ is a
parameter required to make the absorption- and emission-line
metallicities consistent with each other.  M13 report an intrinsic
scatter of 0.38 dex in [M/H] in this relation for DLAs at $0<z<5$.

The aim of this paper is to carry out a critical test of this
statistically derived mass-redshift-metallicity relation (MzZ relation
hereafter) via comparison to individual direct measurements. To do
this, we derive stellar masses from conventional spectral energy
distribution (SED) fits to multi-band photometry of the DLA galaxies.

The structure of the paper is as follows. In Section~\ref{sect:sample}
we describe the sample of confirmed DLA galaxies included in this
investigation and present the photometry for these objects. Based on
the photometry we fit SED models in Section~\ref{sect:SEDfits}. In
Section~\ref{sect:results} we compare the measured stellar masses with
predictions. We discuss the results in Section~\ref{sect:discussion}
and present our conclusions in Section~\ref{sect:conclusions}.

\section{Sample selection}
\label{sect:sample}
Only three DLA galaxies have measured stellar masses reported in the
literature: Q0302--223 \citep{peroux11b}, Q0918+1636 \citep{fynbo11},
and Q2222--0946 \citep{fynbo10,krogager13}. To build a sample of
DLA galaxies with measured stellar masses we included those three and
searched the literature for objects for which the following criteria
were fulfilled:

\begin{enumerate}
\item We select absorbers at cosmological redshifts ($z>0.1$), and
  exclude local galaxy discs that are intersected by QSOs
  \citep[e.g.][]{schulte-ladbeck05}.

\item Absorber-galaxy pair must have a spectroscopically confirmed DLA
  galaxy at an impact parameter smaller than 50~kpc. Choosing a larger
  distance would imply an increasing fraction of interlopers.  Note
  that while this means that we could include DLA galaxies detected at
  zero impact parameter \citep[e.g.][]{noterdaeme12}, they are
  unlikely to pass condition (iii)  because their stellar
  continuum emission is impossible to disentangle from the much
  brighter QSO emission.

\item DLA galaxy must have multi-band photometric measurements in a
  sufficient number of bands to allow conventional SED fitting accurate
  enough to determine the stellar mass of the galaxy. Although more DLA
  galaxies are confirmed than included here
  \citep[e.g.][]{bouche12,peroux14}, the lack of continuum flux
  measurements prevent the derivation of the stellar masses.

\item DLAs must have measured absorption line metallicities,
  preferably from an element that is not depleted onto dust
  grains. Effectively we apply the same criteria which were defined in
  \citet{ledoux06}. Also this requirement de-selects many confirmed
  DLA galaxies at $z<1$ because their corresponding DLA metal
  abundances have not been measured \citep[e.g.][]{chun10}.

\item DLA \ion{H}{i} column density must be high enough that the
  obtained metallicity is reliable, i.e. that a potential ionisation
  correction is not so large as to introduce a dominating systematic
  error. For this we set the following requirement:
  $\log~N$(\ion{H}{i})~$>$~19.5 cm$^{-2}$. Initially we do not make
  any strict distinction between DLAs and sub-DLAs ($19.5 < \log
  N$(\ion{H}{i})$ < 20.3$).

\end{enumerate}

Based on these selection criteria we have collected the sample of DLA
galaxies presented in Table~\ref{tab:DLAdata}.

\begin{table*}
\begin{tabular}{lllllll}
\hline
\hline
QSO name & $z_{\mathrm{DLA}}$ & log N(\ion{H}{i}) & $b$ &
[Zn/H]$_{\mathrm{abs}}$ & [M/H]$_{\mathrm{em}}$ & references\\
          & & [cm$^{-2}$] & [kpc] &  \\
(1)  & (2) & (3) & (4)  & (5)  & (6) &(7)\\
\hline
Q0235+164 ID2 &   0.5243 &  21.70          & 13.2 & --0.60$\pm$0.41 & ~~--0.19$\pm$0.15$^b$  & [1,2,5]\\ 
Q0302--223    &	  1.009  &  20.36$\pm$0.11 & 25.0 & --0.51$\pm$0.12 & $<$0.22$^a$& [1,3,4]\\ 
Q0439--433$^b$    &   0.101  &  19.85$\pm$0.10 & ~~7.2& --0.20$\pm$0.30 & ~~~~0.33$\pm$0.14$^a$ & [1,5]\\
Q0738+313     &   0.2212 &  20.90$\pm$0.07 & 20.3 & --0.70$\pm$0.16 & ~~~~~~~~...& [21,22]\\
Q0827+243     &   0.5247 &  20.30$\pm$0.04 & 38.4 & --0.49$\pm$0.30 & $>$0.01$^b$& [1,5,6,21,22]\\
Q1009--0026   &   0.887  &  19.48$\pm$0.05 & 39.0 & ~~0.25$\pm$0.06 & ~~~~0.31$\pm$0.20$^a$ & [7]\\ 
Q1127--145    &   0.3127 &  21.71$\pm$0.07 & 17.5 & --0.90$\pm$0.11 & ~~~~~~~~...& [2,8,9]\\ 
\hline
Q0528-250     &	  2.811  &  21.35 & ~9.2$\pm$0.2 & --0.91$\pm$0.07  & ~~~~~~~~~...& [10,11,20,23]\\
Q0918+1636    &   2.5832 &  20.96$\pm$0.05 & 16.2$\pm$0.2& --0.12$\pm$0.05 & ~~~0.01$\pm$0.20$^b$  & [12,13,14]\\
Q2206--1958 (total)& 1.921  &  20.65 & 12.7$\pm$0.6   & --0.54$\pm$0.05 & ~~~~~~~~... & [15,16,23]\\
Q2222--0946    &	  2.354  & 20.65$\pm$0.05 & ~~6.3$\pm$0.8 & --0.46$\pm$0.07 &~~--0.44$\pm$0.19$^a$ & [17,18]\\
Q2233+131     &	  3.1501 &  20.00 & 17.9        & --0.80$\pm$0.24  & ~~~~~~~~...  &  [16,19,20]\\
\hline
\end{tabular}
\caption{The sample of DLAs included in this paper. Column 2 gives the
  DLA redshift, Col. 3 the column density of neutral Hydrogen, Col. 4
  the impact parameter as the projected distance between the QSO line
  of sight and the identified DLA galaxy, and Col. 5 the measured
  metallicities.  All absorption metallicities are based on [Zn/H]
  apart from the absorber towards Q2233+131, which converted from
  [Fe/H]+0.3 = [Zn/H] to correct for dust depletion following
  \citep{rafelski12}.  Emission line metallicities in Col. 6 have been
  converted to the same strong-line diagnostics scale ($R_{23}$) as
  described in Sect~\ref{sect:cfunct}.  $^a$ Oxygen abundance
  originally derived via the N2=log([\ion{N}{ii}]/\halpha)
  diagnostics. $^b$ Oxygen abundance originally derived via the
  $R_{23}$ diagnostics.  Apart from emission metallicity conversions,
  all values are adopted from the literature as listed in the
  references in Col. 7.  References: [1] \citet{chen03}, [2]
  \citet{rao11}, [3] \citet{lebrun97}, [4] \citet{peroux11b}, [5]
  \citet{chen05}, [6] \citet{rao03}, [7] \citet{peroux11a}, [8]
  \citet{lane98}, [9] \citet{kacprzak10}, [10] \citet{moller02}, [11]
  \citet{centurion03}, [12] \citet{fynbo11}, [13] \citet{krogager12},
        [14] \citet{fynbo13}, [15] \citet{ledoux06}, [16]
        \citet{weatherley05}, [17] \citet{fynbo10}, [18]
        \citet{krogager13}, [19] \citet{djorgovski96}, [20]
        \citet{moller02}, [21] \citet{kulkarni05}, [22] \citet{rao00},
              [23] \citet{prochaska03}.  }
\label{tab:DLAdata}
\end{table*}

\subsection{Individual DLAs}
\label{sect:individual}
To derive total stellar masses for the galaxies, the magnitudes have
to be extracted from within the same aperture sizes. This is complex
when compiling values from various papers by different authors, in
particular as aperture choices may be very different in \emph{HST} and
ground-based images.

In this section we describe the individual DLAs and their associated
galaxies included in this project. When relevant we apply aperture
corrections appropriate for the given filters. All reported magnitudes
are in the AB system, and when the original magnitudes were given in
the Vega system we included an offset between the two.  In three cases
(Q0235+164, Q0302--223, and Q0738+313) the reported magnitudes included a
correction for Galactic extinction. For the remaining galaxies, we
included the extinction correction as a part of the SED fitting
procedure described in Section~\ref{sect:SEDfits}.

\begin{table*}
\begin{minipage}[t]{0.52\linewidth}
\begin{tabular}{@{}lc@{}ccc}
\hline
\hline
Object & Redshift & Filter & Magnitude  & Reference \\
\hline
DLA 0235+164  & 0.5243 & $U$  &  21.06$\pm$0.03  & [1]$^a$ \\
ID2 & & $B$  &  20.84$\pm$0.01  & [1]$^a$ \\
    & & $V$  &  20.61$\pm$0.01  & [1]$^a$ \\
    & & $R$  &  20.42$\pm$0.01  & [1]$^a$ \\
    & & $I$  &  20.22$\pm$0.02  & [1]$^a$ \\ 
\hline
DLA 0302--223 & 1.009 & WFPC2/F450W &  24.01$\pm$0.34 & [1]$^a$  \\
              &       & WFPC2/F702W &  23.25$\pm$0.06 & [1]$^a$  \\ 
              &       & $I$   &  22.67$\pm$0.04 & [1]$^a$  \\ 
              &       & $J$   &  22.62$\pm$0.17 & [1]$^a$  \\ 
              &       & $H$   &  22.68$\pm$0.25 & [1]$^a$  \\ 
\hline
DLA 0439--433 & 0.101 & $U$  &  19.31$\pm$0.01  & [1]$^a$ \\
& & $B$  &  18.37$\pm$0.01  &  [1]$^a$ \\
& & $V$  &  17.72$\pm$0.01  &  [1]$^a$ \\
& & $I$  &  17.12$\pm$0.01  &  [1]$^a$ \\
& & $J$  &  16.55$\pm$0.01  & [1]$^a$ \\ 
& & $K$  &  16.66$\pm$0.01  & [1]$^a$ \\ 
\hline
DLA 0528--250 & 2.811 & STIS/m$_{50}$ &  25.43$\pm$0.11 & [2] \\
& & WFPC2/F450W &  25.51$\pm$0.10 & [3]$^b$ \\ 
& & WFPC2/F467M &  25.47$\pm$0.13  & [3]$^b$ \\
& & WFPC2/F814W &  24.85$\pm$0.15  & [3] \\
& & NIC2/F160W  &  25.18$\pm$0.22  & [2] \\
\hline
DLA 0738+313 & 0.2212 & $U$  &  23.43$\pm$0.16 & [4]$^a$ \\
& & $B$  &  22.01$\pm$0.05 & [4]$^a$ \\
& & $R$  &  21.09$\pm$0.05 & [4]$^a$ \\
& & $I$  &  20.85$\pm$0.1 & [4]$^a$ \\
& & $J$  &  19.84$\pm$0.1 & [4]$^a$ \\
& & $H$  &  20.09$\pm$0.1 & [4]$^a$ \\
& & $K$  &  19.70$\pm$0.1 & [4]$^a$ \\
\hline
DLA 0827+243 & 0.5247 & $U$  &  23.52$\pm$0.16 & [5] \\
& & $B$  &  22.69$\pm$0.04 & [5] \\
& & $R$  &  21.03$\pm$0.05 & [5] \\
& & $I$  &  20.79$\pm$0.05 & [5] \\
& & $K$  &  18.99$\pm$0.05 & [5] \\
\hline
\end{tabular}
\end{minipage}%
\begin{minipage}[t]{0.52\linewidth}
\begin{tabular}{@{}lc@{}ccc}
\hline
\hline
Object & Redshift & Filter & Magnitude  & Reference \\
\hline
DLA 0918+1636 & 2.583 & NOT/ALFOSC/$u$ & $>$26.5  &  [14] \\
               &       & NOT/ALFOSC/$g$ & 25.9$\pm$0.3  &  [14] \\   
               &       & WFC3/F606W     & 25.46$\pm$0.13  &  [14] \\   
               &       & WFC3/F105W     & 24.61$\pm$0.09  &  [14] \\   
               &       & WFC3/F160W     & 23.63$\pm$0.06  &  [14] \\   
               &       & NOT/NOTCam/\emph{Ks}& $>$23.3   & [14] \\
\hline
DLA 1009--0026 & 0.8870 & SDSS/$u$  &  24.05$\pm$0.79 & [6] \\
& & SDSS/$g$  &  22.95$\pm$0.19 &  [6] \\
& & SDSS/$r$  &  22.28$\pm$0.15 &  [6]  \\
& & SDSS/$i$  &  21.15$\pm$0.08 &  [6]  \\
& & SDSS/$z$  &  20.89$\pm$0.24 &  [6]  \\
& & $J$  &  20.00$\pm$0.04 &  [7]  \\
& & $H$  &  19.78$\pm$0.04 &  [7]  \\
& & $K$  &  19.32$\pm$0.03 &  [7]  \\
\hline
DLA 1127--145 & 0.3127 & $U$  &  23.81$\pm$0.19 & [5] \\
& & $B$          &  24.02$\pm$0.15 & [5]\\
& & $R$          &  21.79$\pm$0.08 & [5] \\
& & WFPC2/F814W  &  22.00$\pm$0.37 & [8] \\
& & $J$          &  22.81$\pm$0.13 & [5] \\
& & $K$          &  22.54$\pm$0.15 & [5] \\
\hline
DLA 2206--1958   & 1.9210 & STIS/m$_{50}$ &  24.10$\pm$0.03 & [2] \\
(source ID2)    &        & NIC2/F160W    &  23.87 & [12] \\
DLA 2206-1958    & 1.9210 & STIS/m$_{50}$ &  23.48 & [13] \\
(total)         &        & WFC3/F160W    & 23.14 & [13] \\  
\hline
DLA 2222--0946  & 2.3535 & WFC3/F606W & 24.29$\pm$0.04 & [15] \\
                &        & WFC3/F105W & 24.51$\pm$0.21 & [15] \\
                &        & WFC3/F160W & 23.53$\pm$0.13 & [15] \\
\hline 
DLA 2233+131 & 3.1501 & $U_n$    &  27.12        & [9] \\
& & $G$          &  25.79 	     & [9] \\
& & STIS/m$_{50}$ & 25.09$\pm$0.2  & [2]$^c$  \\
& & Cousins~$V$  &  25.10$\pm$0.2  & [10] \\
& & $R$          &  25.15        & [9] \\
& & Cousins~$R$  &  24.80$\pm$0.2  & [10] \\
& & WFPC2/F702W  &  24.80$\pm$0.1  & [11] \\
& & NIC2/F160W   &  24.39$\pm$0.2  & [12]$^c$\\
\hline
\end{tabular}
\end{minipage}
\caption{Photometric data of the DLA host galaxy sample. $^a$ Magnitude is corrected for Galactic extinction. $^b$ The magnitude is corrected for the flux contribution from \lya\ as described in the text. $^c$ Magnitude scaled from the given reference.
References: 
[1] \citet{chen03},
[2] \citet{moller02},
[3] \citet{moller98},
[4] \citet{turnshek01},
[5] \citet{rao03},
[6] SDSS-DR10,
[7] \citet{rao11},
[8] \citet{kacprzak10},
[9] \citet{steidel95},
[10] \citet{djorgovski96},
[11] \citet{christensen04},
[12] \citet{warren01},
[13] This paper,
[14] \citet{fynbo13},
[15] \citet{krogager13}.
}

\label{tab:allphot}
\end{table*}

\subsubsection{Q0235+164}
Multiple galaxies are detected within close projected distance from
the QSO line of sight \citep{guillemin97,chen03,rao11}. Three galaxies
within 6\farcs5 corresponding to 40 kpc, are found to have the same
redshift as the DLA \citep{yanny89}. The object at the closest impact
parameter of 1\farcs1 is not resolved in ground based images
\citep[ID1 in][]{chen03}. Due to blending with the background QSO
\citet{chen03} report an upper limit for the galaxy magnitude
WFPC2/F702W $>$21.36. This galaxy is $\sim$1 mag fainter than the
neighbour galaxy at an impact parameter of 2\farcs1 \citep[ID2
  in][]{chen03} which is presumably the dominant component in an
interacting galaxy pair. As in \citet{chen03} we adopt this dominant
galaxy as the primary absorber and reproduce the photometry in
Table~\ref{tab:allphot}.

\subsubsection{Q0302--223}
Candidate DLA galaxies nearby in projection to the QSO were detected
by \citet{lebrun97}, and multi-band photometry of these galaxies
reported by \citet{chen03}. Objects named 3+4 by these authors were
spectroscopically confirmed to be at the DLA redshift by
\citet{peroux11b}. The photometry of the sum of the two galaxy
components is listed in Table~\ref{tab:allphot}.

\subsubsection{Q0439--433}
This QSO has the lowest redshift DLA in our sample ($z=0.101$). The
galaxy responsible for the absorption line was identified by
\citet{chen03}, and we reproduce the photometry of the galaxy in
Table~\ref{tab:allphot}.

\subsubsection{PKS 0528--250}
Extensive \emph{HST} data exist for this QSO, and the DLA galaxy
magnitudes are adopted from \citet{moller98,warren01,moller02}. The
STIS magnitude is based on the best fit S\'ersic model profile
extrapolated to infinite radius, while the WFPC2 magnitudes are found
within a 1\arcsec\ diameter aperture, which should provide the total
magnitudes of the galaxy given the zero-point adopted by
\citet{moller98}. Two of the filters (WFPC2/F467M and WFPC2/F450W) are
contaminated by \lya\ emission from the DLA galaxy. The extended
\lya\ emitting region has a flux of $(7.4\pm0.6)\times10^{-17}$
\ecs\ \citep{warren96}. However, compared to the extension of the
\lya\ emission, the continuum emission from the galaxy is considerably
more compact. We estimate that the contamination from the
\lya\ emission line in the two bands amounts to $1.4\times10^{-17}$
\ecs\ by assuming that the total flux in each band is the sum of the
continuum flux density multiplied by the filter width plus the
\lya\ flux. We correct the broad band magnitudes by subtracting the
contributions of the \lya\ line flux. Since the filter width of the
STIS/m$_{50}$ is very large, the \lya\ emission does not contribute
significantly to its magnitude in this band (0.04 mag), whereas the
correction for the intermediate-width filter F467M data is 0.31 mag.

For the NICMOS data \citet{warren01} report a magnitude of 25.54
within a 0\farcs45 diameter aperture, while \citet{moller02} report
$25.18\pm0.22$ mag by extrapolating the model to infinite radius. We
summarise the photometry in Table~\ref{tab:allphot}.

\subsubsection{Q0738+313}
The galaxy responsible for the DLA towards Q0738+313 was identified by
\citet{turnshek01}. The photometry of the galaxy is reproduced in
Table~\ref{tab:allphot}, here converted to AB magnitudes by adding the
offsets between Vega and AB magnitudes of $0.63, -0.07, 0.21, 0.45,
0.94, 1.39$, and $1.90$ to the magnitudes measured in the $UBRIJH$ and
$K$ filters, respectively.

\subsubsection{Q0827+243}
The galaxy responsible for the DLA towards Q0827+243 was identified by
\citet{rao03}. The photometry of the galaxy is reproduced in
Table~\ref{tab:allphot}, here converted to AB magnitudes.

\subsubsection{Q0918+1636}
The DLA galaxy was discovered in spectroscopic observations by \citet{fynbo11},
and followed-up with \emph{HST} and ground-based imaging in \citet{fynbo13}.
The photometry from \citet{fynbo13} is reproduced in Table~\ref{tab:allphot}.

\subsubsection{Q1009--0026}
The sub-DLA galaxy towards Q1009--0026 was identified by
\citet{peroux11a} at a rather large impact parameter of 39 kpc. The
optical photometry of the galaxy is adopted from the SDSS DR10
database and the near-IR from \citet{rao11} as listed in
Table~\ref{tab:allphot}.

\subsubsection{PKS 1127--145}
The field around PKS 1127--145 is crowded, and a galaxy at an impact
parameter of 9\farcs6 (corresponding to 44~kpc) at the same redshift
as the DLA was originally identified by \citet{bergeron91}. Later
observations showed that also galaxies closer to the QSO line of sight
lie at the DLA redshift \citep{rao03,chen03}. At least five galaxies
within 17--240~kpc in projection have been identified at the DLA
redshift \citep{kacprzak10}, revealing the origin of the DLA in a
group environment. The photometry of the galaxy detected at the
smallest impact parameter is reproduced in Table~\ref{tab:allphot}.

\subsubsection{Q2206--1958}
The morphology of the DLA galaxy is very complex with 3 compact
sources and extended emission covering a size of roughly
2\arcsec$\times$1\arcsec\ \citep{moller02}. The NICMOS data presented
in \citet{warren01} show no extended emission because of its
insufficient depth. We know from spectroscopic measurements that the
compact knots arise at the same redshifts
\citep{moller02,weatherley05}, and therefore the whole region has a
complex morphology, which makes it likely that the DLA arises in a
compact, merging galaxy group.

Photometry of source ID1 \citep[named N-14-1C in][]{warren01} is
STIS/m$_{50}~=~24.10\pm0.03$ within a 0\farcs9 diameter aperture
\citep{moller02}.  Within a 0\farcs45 diameter aperture
\citet{warren01} report NIC2/F160W~=~25.11. An aperture correction to
a 0\farcs9 diameter of 0.66 mag gives NIC2/F160W$~=~24.45$ mag.  For
the source ID2 \citep[named N-14-2C in][]{warren01} the corresponding
numbers is NIC2/F160W$~=~23.87$ mag within a 0\farcs9 diameter
aperture.
To verify the NICMOS photometry, we use archive \emph{HST}/WFC3 F160W
data (proposal ID: 11694, PI: Law). With a total integration time of
8093~s the data is substantially deeper than the NICMOS data. Due to
the depth of the WFC3 data, we confirm that the extended emission
visible in the STIS data is also present in the observed near-IR
wavelengths. Due to PSF subtraction residuals we do not recover the
emission from the compact source ID1.  For the compact source ID2,we
derive WFC3/F160W~=~24.22 mag within an aperture size of 0\farcs9.

Because the morphology of the DLA galaxy is complex, we present both
the flux from object ID2, plus the total integrated flux from the
entire complex region in Table~\ref{tab:allphot}.  Re-examining the
STIS data of the DLA host from \citet{moller02} we derive an
integrated magnitude of STIS/50CCD = 23.48 mag while the WFC3/F160W
data in the same aperture gives 23.14 mag.

From the measured flux and impact parameters for the various sources
reported in \citet{warren01}, we calculate a flux-weighted mean impact
parameter of 1\farcs48$\pm0.06$, corresponding to $12.7\pm0.6$ kpc at
the DLA redshift. In the remainder of the paper, we adopt these values
and the total integrated flux for the analysis of the DLA galaxy.

\subsubsection{Q2222--0946}
The DLA galaxy was discovered by \citet{fynbo10}, and deeper data
including {\it HST} images of the stellar light of the host galaxy was
presented in \citet{krogager13}. The DLA galaxy abundances and
kinematics are further investigated from integral field spectra
\citep{peroux12,jorgenson14}.  The photometry of the galaxy is
reproduced in Table~\ref{tab:allphot}.

\subsubsection{Q2233+131}

The DLA galaxy was discovered in \citet{steidel95}. Within a 0\farcs9
diameter aperture the measured magnitudes are STIS/m$_{50}$~=~25.75
mag \citep{moller02} and NIC2/F160W~=~25.05 \citep{warren01}.  These
values are faint compared to ground based $V$ and $R$ band data. The
aperture correction to total magnitudes for compact objects in
NIC2/F160W imaging data is 0.66 mag \citep{warren01}.  Assuming that
the colour of the galaxy is constant with radius this aperture
correction is also added to the STIS/m$_{50}$ magnitude.  This gives
the photometry data listed in Table~\ref{tab:allphot}.

While \lya\ emission is also detected for the DLA galaxy towards
Q2233+131, the emission falls at a wavelength at the edges of the $V$
band filters where the transmission is low. We therefore do not
correct the measured magnitudes for contamination by \lya.


\begin{figure*}
\begin{center}
\includegraphics[width=15.cm, bb=45 70 513 770,clip]{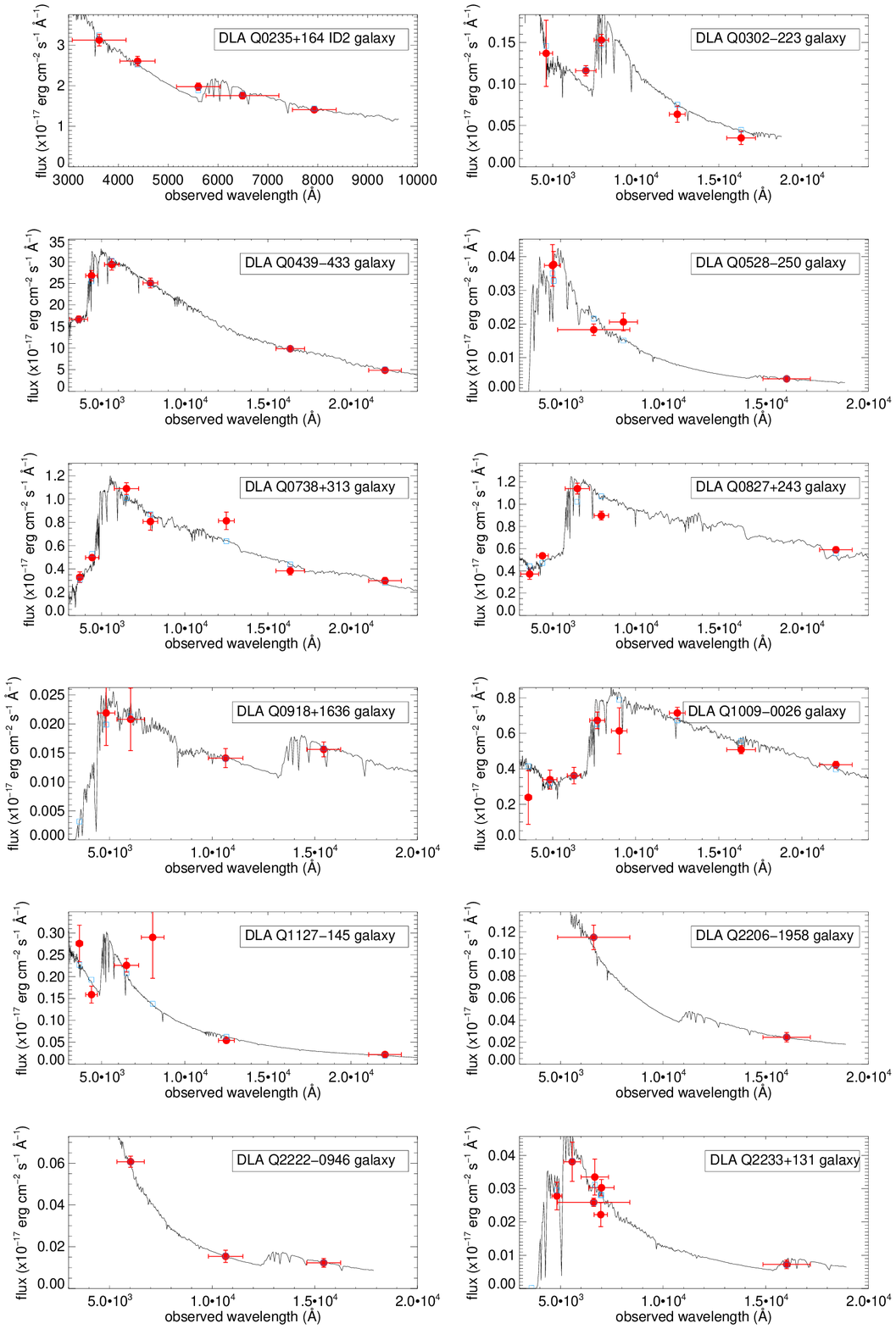}
\end{center}
\caption{Best fit SED models for all the DLA galaxies. The red dots
  denote the measured photometric points and the blue open squares are
  the expected flux density from the best fit SED model. \textit{See
    the online version of the journal for a colour version of this
    figure.}}
\label{fig:seds}
\end{figure*}

\section{SED fitting}
\label{sect:SEDfits}
In this paper we are primarily interested in deriving the stellar
masses for the galaxies, and adopt a standard method of fitting
spectral energy distribution models to the measured photometry
provided in Section~\ref{sect:individual}. We use the release version
12.2 of {\sc HyperZ} \citep{bolzonella00} fixing the redshift to that
of the absorber.  Luminosity distances and impact parameters in kpc
are calculated assuming a flat cosmology with
$H_0=70.4$~km~s$^{-1}$~Mpc$^{-1}$ and $\Omega_{\Lambda}~=~0.727$
\citep{komatsu11}.

We retrieve the relevant filter transmission curves from the various
observatories and instruments used to measure the original galaxy
magnitudes. We assume a 0.1 magnitude uncertainty when those are not
stated directly in the literature.  Three of the galaxies listed in
Table~\ref{tab:allphot} have already had their magnitudes corrected
for Galactic extinction, while the remaining galaxy magnitudes are
corrected for Galactic extinction from the maps calibrated by
\citet{schlafly11} as part of the SED fitting procedure. Intrinsic
reddening is included as a free parameter to the SED fits and allowed
to vary between $A_V=0-1$ magnitude.

Input spectral templates are extracted from \citet{bruzual03} models,
derived with Padova 1994 stellar evolutionary tracks with a
\citet{chabrier03} IMF, and metallicities of 0.2, 0.4 and 1.0 times
solar.  Various star formation histories are assumed including a
single instantaneous burst, a constant star formation rate, and
exponentially decreasing SFRs with time scales, $\tau$ between 10 Myr
and 1 Gyr  as well as exponentially increasing SFRs with $\tau$
  between 10 Myr and 500 Myr.  At high redshifts, absorption by
neutral hydrogen in the Lyman alpha forest is taken into account using
the description of the opacity in \citet{madau95}, see also
\citet{moller90}.

The stellar masses of the galaxies are estimated from the best fit
template from {\sc HyperZ}.  The best fit model templates are in
  all cases instantaneous burst models or exponentially decreasing
  models with $\tau$=10--100 Myr. The data and the best fit template
spectra are illustrated in Fig.~\ref{fig:seds}. To propagate the
magnitude errors and obtain uncertainties on the stellar masses we
assume a Gaussian error distribution, and then perform SED fits to
$10^3$ random realisations for each galaxy. While we derive the
stellar masses based on the best fit template spectrum, other star
formation histories can also yield acceptable SED fits. We examine the
differences for the various star formation histories, and determine
that for acceptable fits the stellar masses are consistent to within
0.2 dex.

The \citet{bruzual03} spectral templates do not include nebular
emission lines, and SED fits for galaxies with very strong emission
lines relative to their continuum emission reveal much lower masses
and high specific star formation rates \citep{schaerer09,watson11}.
However, since DLA galaxies with extremely high specific star
formation rates are not yet discovered, we do not expect a significant
effect. Nevertheless, we examine the effect emission lines would have
for the derived stellar masses. Since {\sc HyperZ} does not include
emission lines in the template spectra, we use another photometric
redshift code, {\sc Le Phare} \citep{ilbert09} to investigate the
effect from nebular emission lines on the best fit spectral
templates. Including emission lines, the best fit stellar mass
decrease by 0.01--0.05 dex only, with the most extreme case being
Q2233+131, where the stellar mass is 0.16~dex smaller.

Stellar masses obtained with a Chabrier IMF have been published for
the DLA galaxies towards Q2222--0946, Q0918+1636, and Q0302--223.
Because of possible offsets between masses obtained from various SED
fitting codes, we derive the stellar masses from {\sc HyperZ}. A different
stellar mass can be explained by the various codes finding different
ages, extinctions and star formation histories for the best
fits. Besides, the choice of stellar evolutionary tracks for creating
the input spectral templates also result in a different stellar mass.
As listed in Table~\ref{tab:gradients} we find
$\log(M_*/M_{\odot})=9.65\pm0.08$ compared to
$\log(M_*/M_{\odot})=9.5$ \citep{peroux11b} for the DLA galaxy towards
Q0302--223.  For the DLA galaxy towards Q0918+1636 we derive stellar
masses of $\log(M_*/M_{\odot})=10.33\pm0.18$ compared to
$\log(M_*/M_{\odot})=10.10\pm0.14$ \citep{fynbo13}, and for
Q2222--0946, we find $\log(M_*/M_{\odot})=9.62\pm0.12$ compared to
$\log(M_*/M_{\odot})=9.32\pm0.23$ \citep{krogager13}. The masses are
all consistent to within about $1\sigma$ uncertainties. For
consistency of the further analysis we use only stellar masses derived
using {\sc HyperZ}.

\begin{table*}
\begin{tabular}{lllllrrr}
\hline
\hline
QSO name & $z_{\mathrm{DLA}}$ & 
log~$M_*$($C_{\rm[M/H]}=0.44)$&  log~$M_* ({\mathrm{SED})}$ &
log~$M_*^{\mathrm{DLA}}$   & $C_{\rm[M/H]}$    &  $\Gamma$ \\
          & &  [M$_{\odot}$] & [M$_{\odot}$] &  [M$_{\odot}$] &  &[dex kpc$^{-1}$] & \\
(1)  & (2) & (3) & (4)  & (5)  & (6) & (7)\\
\hline
Q0235+164 ID2 &   0.5243 & ~~8.91$\pm$0.72 &  ~~9.66$\pm$0.01  & ~~8.65$\pm$0.72  & 0.87$\pm$0.41 & 0.031$\pm$0.033 \\ 
Q0302--223    &	  1.009  & ~~9.37$\pm$0.21 &  ~~9.65$\pm$0.08  & ~~9.56$\pm$0.21  & 0.60$\pm$0.13 & $<$0.029\\
Q0439--433    &   0.101  & ~~9.36$\pm$0.53 & 10.01$\pm$0.02    & ~~8.86$\pm$0.53  & 0.81$\pm$0.30 & 0.074$\pm$0.046 \\
Q0738+313     &   0.2212 & ~~8.55$\pm$0.28 & ~~9.33$\pm$0.05   & ~~8.56$\pm$0.28  & 0.88$\pm$0.16 & \\
Q0827+243     &   0.5247 & ~~9.11$\pm$0.53 & 10.09$\pm$0.15    & ~~9.82$\pm$0.53  & 1.00$\pm$0.31 & $>$0.013 \\
Q1009--0026   &   0.887  &  10.63$\pm$0.11 & 11.06$\pm$0.03    &  11.37$\pm$0.11   & 0.68$\pm$0.06 & 0.002$\pm$0.005 \\
Q1127--145    &   0.3127 & ~~8.25$\pm$0.19 & ~~8.29$\pm$0.09   & ~~8.16$\pm$0.19  & 0.46$\pm$0.12 \\
\hline
Q0528-250     &	  2.811  & ~~9.64$\pm$0.12 &  ~~8.79$\pm$0.15  & ~~9.22$\pm$0.13   & --0.05$\pm$0.14 \\
Q0918+1636    &   2.5832 &  11.02$\pm$0.09 &   10.33$\pm$0.08  &  10.87$\pm$0.10   & 0.05$\pm$0.07  & 0.008$\pm$0.013\\
Q2206-1958 (total)& 1.921& ~~9.88$\pm$0.09 &  ~~9.45$\pm$0.30  & ~~9.60$\pm$0.13   & 0.20$\pm$0.18 \\
Q2222-0946    &	  2.354  & 10.29$\pm$0.12 &  ~~9.62$\pm$0.12   & ~~9.75$\pm$0.14   & 0.06$\pm$0.10 & 0.003$\pm$0.032\\
Q2233+131     &	  3.1501 & ~~9.84$\pm$0.42 &  ~~9.85$\pm$0.14  & ~~9.76$\pm$0.42   & 0.45$\pm$0.25 \\
\hline
\end{tabular}
\caption{Predicted and measured stellar masses for the DLA galaxies.
  Columns 3 --7 present the results of this paper as described in
  Section~\ref{sect:results}. }
\label{tab:gradients}
\end{table*}



\section{Test of the DLA galaxy mass-redshift-metallicity relation}
\label{sect:results}
The primary aim of this paper is to test the statistically derived
MzZ relation reproduced in Eq.~\ref{eq:pmfit}. In the
previous section we determined $M_*$ for 12 DLA galaxies from SED fits
and we now compare those to the values determined from
Eq.~\ref{eq:pmfit}.  This constitutes a critical test.

\subsection{Determination of $C_{\mathrm{[M/H]}}$}

Eq.~\ref{eq:pmfit} contains a term, $C_{\mathrm{[M/H]}}$, which
describes the difference between the gas-phase metallicity measured
via absorption lines and emission lines. Empirically this difference
is expected to be in the range 0 -- 0.5 dex (M13). We can therefore
for each galaxy subtract the expression for log$M_*$ given in
Eq.~\ref{eq:pmfit} from the measured log$M_*\mathrm{(SED)}$ and
solve for $C_{\mathrm{[M/H]}}$.

In Table~\ref{tab:gradients} Col. 6, we list the individual values
determined this way, as well as corresponding errors propagated
from measurement errors on
metallicities and log$M_*\mathrm{(SED)}$.  It is seen that the scatter
of the values is larger than the individual errors, which is caused by
the intrinsic scatter of the relation in Eq.~\ref{eq:pmfit}. In order
to obtain an empirically determined expectation value for
$C_{\mathrm{[M/H]}}$ we adopt the $C^2_{\mathrm{dof}}$ minimisation
method described in M13, i.e. we describe the total scatter as the
square sum of the ``natural intrinsic scatter'' and measurement errors
($\sigma_{\rm tot}(i)^2 = \sigma_{\rm nat}^2 + \sigma_{C}(i)^2$) where
$\sigma_{C}(i)$ is the propagated error on individual values of
$C_{\mathrm{[M/H]}}$. We can now use

\begin{eqnarray}
C^2_{\mathrm{dof}} = \sum_{i=1}^{i=12} & (
( C_{\mathrm{[M/H]}}(i) - \langle C_{\mathrm{[M/H]}} \rangle )
/ \sigma_{\mathrm{tot}}(i))^2 /dof
\label{eq:chimin}
\end{eqnarray}
to obtain the expectation value $\langle C_{\mathrm{[M/H]}}\rangle$
which provides
the smallest scatter and for which $C^2_{\mathrm{dof}} = 1$. The
degrees of freedom is here $dof=11$ (for details on the method see
M13). We find $\langle C_{\mathrm{[M/H]}}\rangle = 0.44 \pm 0.10$ and
the intrinsic scatter of $C_{\mathrm{[M/H]}}$ is $\sigma_{\rm nat} = 0.31$.

\begin{figure}
\begin{center}
\includegraphics[bb=150 365 525 700,clip, width=8.5cm]{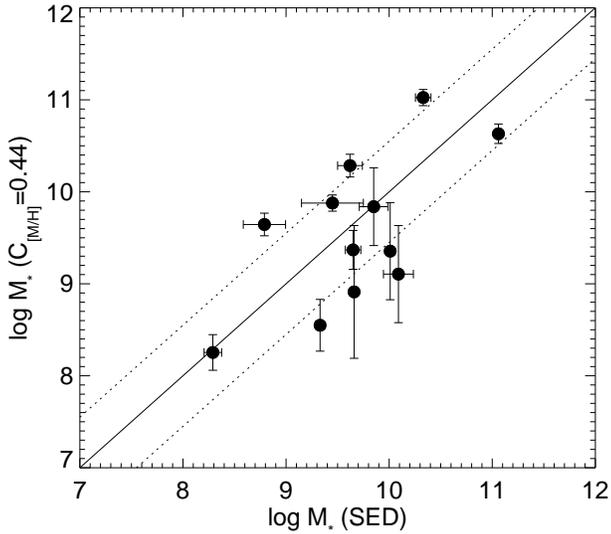}
\end{center}
\caption{Predicted stellar masses of DLA galaxies based on
  Eq.~\ref{eq:pmfit} with $C_{\mathrm{[M/H]}} =0.44$ as a function of
  the stellar mass measured directly from the SED fits. The straight
  solid line represents equal masses and the measured scatter
 of $\pm$0.55 dex in log$M_*$ is represented by the dotted straight lines.
}
\label{fig:mcomp}
\end{figure}

Using $C_{\rm [M/H]} = 0.44$ in Eq.~\ref{eq:pmfit} we then calculate
the values for $M_*$($C_{\rm [M/H]} = 0.44$) listed in Col.~3 in
Table~\ref{tab:gradients}. In Fig.~\ref{fig:mcomp} we plot those
values versus the $M_*$(SED) values determined in
Sect.~\ref{sect:SEDfits}. The dotted lines mark the range of the
intrinsic scatter converted to stellar mass, $\pm 1.76 \times 0.31 =
\pm 0.55$ dex in log$M_*$. It is seen that the SED fits are consistent
with the computed values to within the internal scatter over 3 decades
in stellar mass. We conclude that the MzZ relation in
Eq.~\ref{eq:pmfit} has been confirmed at stellar masses above $10^8$
M$_{\odot}$.

\subsection{Is $C_{\mathrm{[M/H]}}$ a function rather than a constant?}
\label{sect:cfunct}

In the previous section we treated $C_{\rm [M/H]}$ as a constant
offset between metallicities determined in absorption and emission.
We may attempt to go one step further and ask what is the cause of the
non-zero value of $C_{\rm [M/H]}$. An absorption metallicity is
sampling the gas in a very narrow beam along the sight-line to the
background quasar, and this DLA sight-line is typically offset some
distance from the sight-line through the centre of the galaxy. In our
sample this distance (commonly referred to as the ``impact parameter''
$b$) is in the range 6 -- 39 kpc. Emission line metallicities are
measured at the slit position chosen by the observer (usually the
centre of the galaxy) and it represents the mean metallicity taken
across the slit (typically $\pm 3$ kpc at low redshifts, $z\sim 0.5$;
$\pm 4$ kpc at higher redshifts, $z\sim2.5$). Any non-zero value
of $C_{\rm [M/H]}$ therefore represents a variation of
enrichment histories of the gas seen in different sight-lines. The
enrichment could be caused by local ``in situ'' star-formation, but
might also be the result of inflow of metal-rich gas from larger
distances. In case those enrichment histories are randomly scattered
across the galaxy and its halo, we would expect $C_{\rm [M/H]}$ to be
zero in the mean.

We have searched the literature and found that for five of the
galaxies in our sample strong emission-line metallicities have been
reported \citep{chen05,peroux12,krogager13,fynbo13}. Historically
several different strong-line diagnostics have been used \citep[for a
  review see][]{kewley08} so we convert all the reported metallicities
to the same $R_{23}$ diagnostics \citep{kewley02} using the equations
given in \citet{kewley08}.  The choice to use the \citet{kewley02}
diagnostics is crucial because the calibration of Eq.~\ref{eq:pmfit}
is tied to the MzZ relation in \citet{maiolino08} who used exactly
this metallicity diagnostics. I.e. we use internally consistent
calibrations of the metallicities.  Col. 6 in Table~\ref{tab:DLAdata}
lists emission metallicities, [M/H]$_{\mathrm{em}}$, relative to a
solar value of 12+log(O/H)~=~8.69 \citep{asplund09}.

From Cols. 5 and 6 in Table~\ref{tab:DLAdata} it is seen that the
measured difference between the emission and absorption metallicities
in all five cases is positive. In Table~\ref{tab:DLAdata} we also
report one lower limit (where the difference is also positive), and an
upper limit (where the difference could be either positive or
negative).  The mean of the five values is $C_{\rm [M/H]}=0.22\pm0.11$.

With the caveat
that five objects make up a very small sample, it is nevertheless
noteworthy that we consistently find highest metallicities in the
centre. If this is a real effect rather than a result of random
scatter in a small sample, then it would imply that metallicity, in
the mean, could be a function of the impact parameter. In this case
the same result should appear in the entire sample. In the following
section we shall briefly test if this prediction holds.

\subsection{Is $C_{\mathrm{[M/H]}}$ a function of the impact
parameter?}
\label{sect:gradient}

We now test the hypothesis that $C_{\rm [M/H]}$ can be expressed
as a simple linear function of the impact parameter $b$:
$C_{\rm [M/H]} = \Gamma b$.  First we
consider only the five galaxies for which we have both central
metallicity measurements and a second measurement obtained at a
distance $b$ from the centre. The two measurements of [M/H] and the
impact parameter allow us to compute $\Gamma$ directly:
\[\Gamma=(\rm{[M/H]}_{\mathrm{em}}-\rm{[M/H]}_{\mathrm{abs}})/b.\]
Those values (and propagated errors) are listed in Col. 7 in
Table~\ref{tab:gradients}.
We compute the mean and find $\Gamma = 0.023\pm0.015$.

In order to perform a test on the entire sample we
repeat the calculation of $C^2/dof$ from Eq.~\ref{eq:chimin},
but now replacing the constant $\langle C_{\mathrm{[M/H]}} \rangle$
by $\Gamma b(i)$. We vary
$\Gamma$ and again seek the solution for $C^2/dof=1$ with the smallest
internal scatter. We find $\Gamma = 0.022 \pm 0.004$, 
in excellent agreement with the value of $\Gamma$ determined from
emission line and absorption metallicities above. The two
determinations are completely independent and they both indicate that
$C_{\rm [M/H]}$ can be well approximated as $0.022 b$, i.e.
that Eq.~\ref{eq:pmfit} may be expanded to

\begin{equation}
\log(M_*^{\mathrm{DLA}}/M_{\odot})= 1.76(\mathrm{[M/H]} + 0.022b +
0.35z + 5.04).
\label{eq:newfit}
\end{equation}

\begin{figure}
\begin{center}
\includegraphics[bb=150 365 525 700,clip, width=8.5cm]{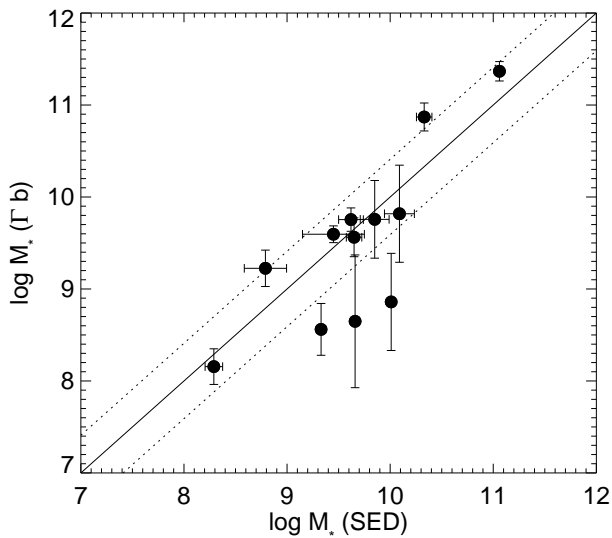}
\end{center}
\caption{As Fig. 2, but the mass prediction is based on
  Eq.~\ref{eq:newfit}. The scatter around this relation is reduced to
  0.39 dex in log$M_*$.  }
\label{fig:mcomp2}
\end{figure}

Eq.~\ref{eq:newfit} is an alternative way to write Eq.~\ref{eq:pmfit},
and it contains the same number of fitted parameters (the constant
$\Gamma$ has replaced the constant $C_{\rm [M/H]}$).  If
Eq.~\ref{eq:newfit} now provides a smaller scatter than before, then
it means that it provides a better fit. Here we test if this is the
case. Using the prescription in Eq.~\ref{eq:newfit} we find an
internal scatter of 0.39 in log $M_*$. Using $C_{\mathrm{[M/H]}}=0.44$
the corresponding scatter was 0.55 so in terms of improvement of the
fit we see that $(0.55^2 - 0.39^2)^{1/2} = 0.39$, which means that by
introducing the impact parameter dependency in Eq.~\ref{eq:newfit} we
may have identified the source of half of the scatter in the original
DLA-$M_*$ prescription.  Values for $\log(M_*^{\mathrm{DLA}})$ are
listed in Col. 5 in Table~\ref{tab:gradients}, and a visual impression
of the smaller scatter is provided in Fig.~\ref{fig:mcomp2} where we
again plot predicted versus measured stellar masses.

M13 report a scatter of $0.38$~dex in [M/H] in Eq.~\ref{eq:pmfit} by
combining DLA absorption metallicities with the MZ
scaling relation used in \citet{maiolino08}. Converting this to a
scatter in stellar mass gives $0.38\times1.76~=~0.67$ dex in
log~$M_*$.  With the caveat that the sample used here is only of 12
galaxies with measured impact parameters, we conclude that a
significant improvement has been achieved over the prescription in
M13.

\subsection{DLA galaxy sample tests}
 Fields around quasars often contain several faint galaxies which
  may be related to either an absorber, to the quasar, or to
  neither. Usually redshifts are available for only a small fraction
  of the galaxies, but it is always a cause to worry if
  mis-identifications might cause false results. For this reason we
  have carried out extensive tests as detailed in the Appendix. We
  demonstrate that the results are robust and that choosing other
  samples increase the scatter around Eqs. 1 and 3, and conclude that
  the chosen sample is very likely to represent the majority of
  correctly identified DLA galaxies.

We test if the range of impact parameters between sub-samples affects
the results. By splitting the sample in two with impact parameters
less than 15 kpc and and larger than 15 kpc, we find
$\Gamma=0.023\pm0.014$ and $\Gamma=0.022\pm0.004$, respectively,
showing that the slope is stable for choice of sub-samples.  

We also briefly investigate the outcome if we exclude the three
sub-DLAs from the analysis.  The resulting internal scatter does not
change significantly and we obtain an insignificant change for the
slope $\Gamma$ to $0.024\pm0.005$ dex~kpc$^{-1}$. We conclude that
there is no evidence in the current sample that sub-DLAs follow a
different MzZ relation from that of DLAs.

\subsection{Comparing with galaxy MZ relations}
With our sample of measured (SED) stellar masses and absorption
metallicities we can now investigate how these values compare to the
MZ relation derived from luminosity-selected galaxy samples. Several
studies have demonstrated a redshift evolution of the MZ relation
\citep[e.g.][]{tremonti04,savaglio05,erb06,maiolino08}. However, as
discussed above, a comparison between MZ relations at different
redshifts is only valid provided metallicities are derived from the
same diagnostics.  As our mass determination is based on the
  calibration by \citet{maiolino08} we can make a reliable
  comparison.

\begin{figure*}
\begin{center}
\includegraphics[bb=150 365 550 700,clip, width=8.5cm]{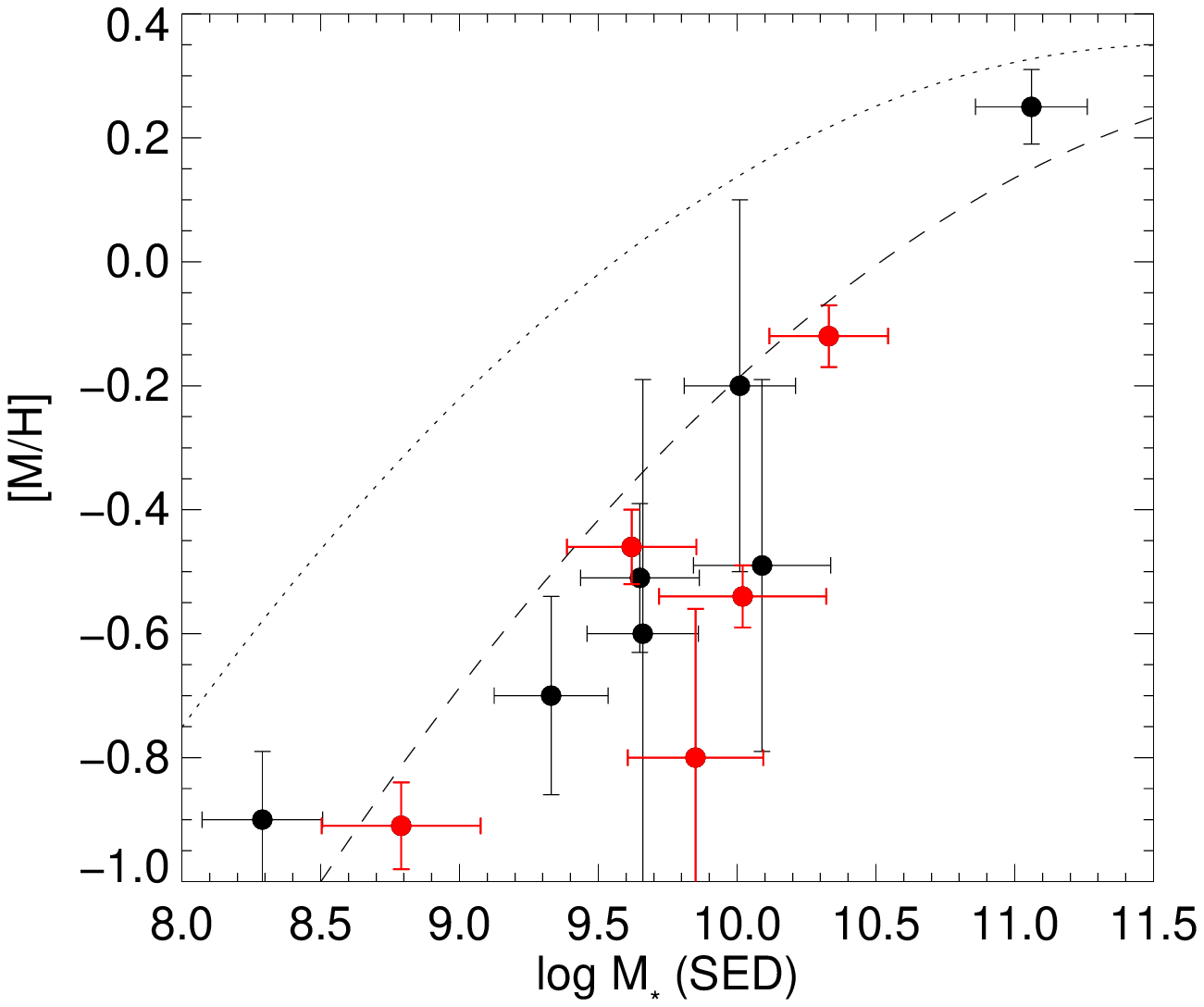}
\includegraphics[bb=150 365 550 700,clip, width=8.5cm]{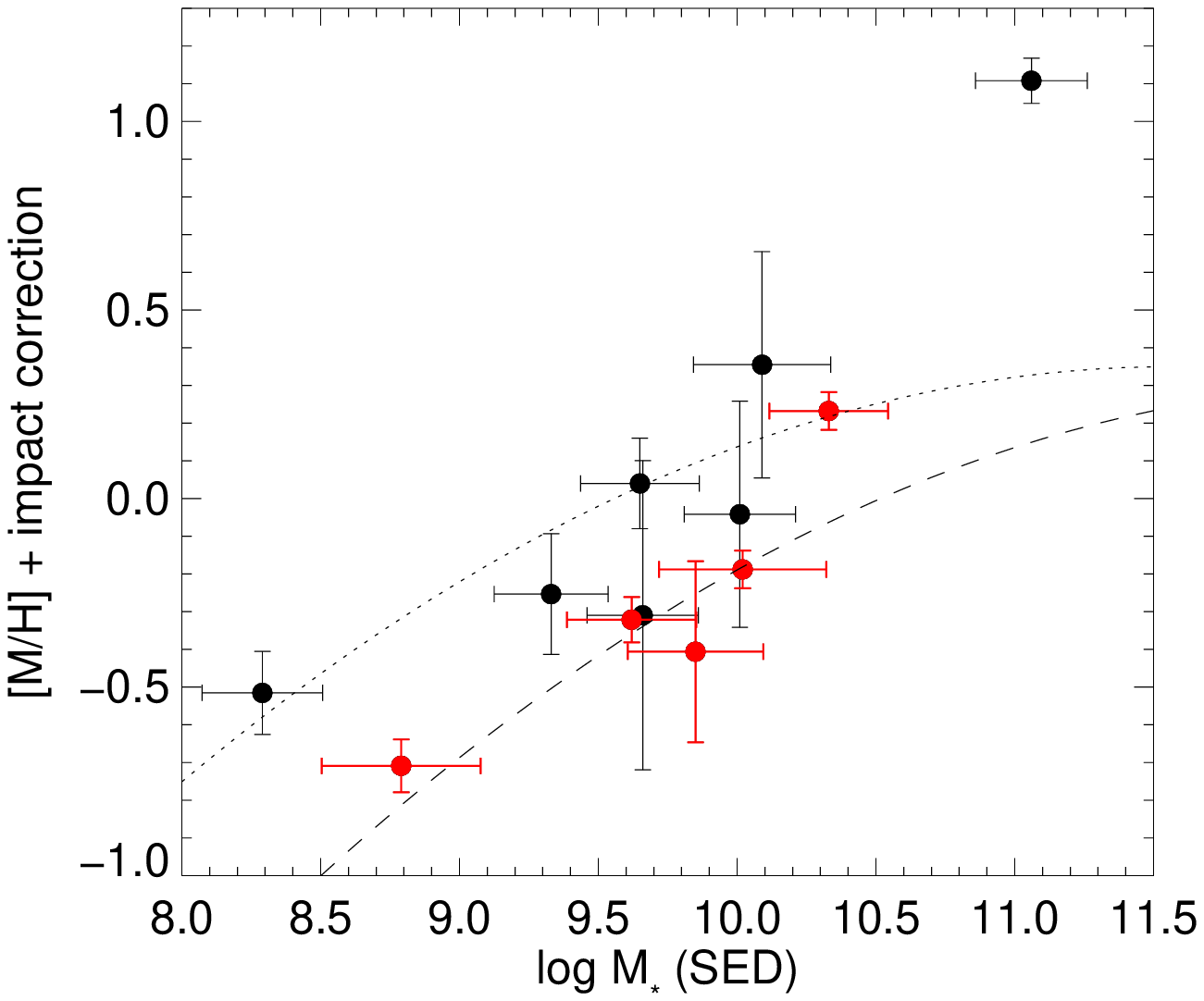}
\end{center}
\caption{\emph{Left panel}: Observed DLA metallicity versus the galaxy
  stellar mass determined from the SED fits. In both panels, the
  dotted and dashed curves represent the observed MZ relation of
  luminosity selected galaxies at $z=0.7$ and $z=2.2$, respectively
  \citep{maiolino08}. \emph{Right panel}: The DLA metallicities have
  been corrected by adding a term, $\Gamma b$, reflecting the average
   metallicity offset ($\Gamma=0.022\pm0.004$)  and measured
  impact parameters $b$. DLAs at $z>2$ (median $z=2.56$) are shown as
  red symbols, and $z<1$ (median $z=0.52$) are shown as black
  symbols. Ten out of 12 targets now follow the redshift-dependent MZ
  relation within 1$\sigma$ measurement uncertainties. }
\label{fig:MZ}
\end{figure*}

Figure~\ref{fig:MZ} (left panel) shows the location of the 12 DLAs in
our sample compared to the $z=0.7$ and $z=2.2$ MZ relations from
\citet{maiolino08}. We see immediately that most of the DLAs lie below
the MZ relation at either redshift.  DLAs at
$z\gtrsim 2$ are colour coded red while those at $z\lesssim 1.0$ are
black. Neither sample follow their respective MZ relation at their
given redshift even when including the average correction to the
metallicity $\langle C_{\mathrm{[M/H]}}\rangle$,
and high- and low-redshift DLAs are mixed in the diagram. In the right
hand panel of Fig.~\ref{fig:MZ} we show the same plot but here we have
added $\Gamma b$, with $\Gamma=0.022$, to the absorption metallicity
of each galaxy thereby correcting them to the assumed metallicity in
the centre. It is striking that not only do the DLA galaxies now
occupy the same general part of the plot as the emission selected
galaxies, they also separate out to follow the same specific redshift
tracks.

\section{Discussion}
\label{sect:discussion}

The average impact parameter for the 7 low redshift DLA galaxies is
$\langle b \rangle = 22.9$ kpc while it is only $12.5$ kpc for the
high redshift DLAs (Table~\ref{tab:DLAdata}, Col. 4). This difference
may reflect a real evolution, but it may also be a result of selection
effects. In either case this anti-correlation between $b$ and $z$ in
our sample means that any correlation with $b$ will have a degenerate
solution as a correlation with $z$. As a result, when we in
Sect.~\ref{sect:gradient} find that $C_{\rm [M/H]}$ can be expanded as
$\Gamma b$, it could equally well be written $\Psi z$, and this
solution would result in a figure very similar to
Fig.~\ref{fig:mcomp2}. The correlation is significant so one of them
must be true, but can we decide which?

We first note that if we assume that the correlation is with redshift
then it almost cancels the redshift dependent term in
Eq.~\ref{eq:pmfit}. This redshift evolution of the MZ relation has
been determined from large samples independent of this work, both
emission selected samples \citep{tremonti04,erb06,maiolino08} and
absorption selected samples
\citep{ledoux06,neeleman13,moller13}. Therefore, the solution as $\Psi
z$ is ruled out by all those independent results.  Second we note that
the existence of metallicity  offsets between the galaxy and QSO
  lines of sight. The value for the metallicity offset per kpc is
  comparable to gradients widely seen in local galaxies
\citep[e.g.][]{zaritsky94} as well as at $0.8<z<2.2$
\citep{swinbank12}.  In addition several metallicity gradients for DLA
galaxies have been reported
\citep{chen05,peroux11a,peroux13,krogager13,fynbo13,peroux14}, but
what the cause is of these metallicity gradients is not known. It
could be variations in local in-situ star formation, but it is
possible that large-scale gas flows are also important. In particular
at distances larger than $\sim$20 kpc in- and outflows of metal-poor,
and metal-rich gas, respectively are likely to play a major role in
the enrichment history of the gas.  E.g. \citet{bouche13} argue that
the lower metallicity of a $z\sim2$ DLA towards HE 2243--60 with the
DLA galaxy detected at an impact parameter 26 kpc can be explained by
dilution of the DLA from pristine material.

By decomposing the term for $C_{\rm [M/H]}=\Gamma b$ in
Eq.~\ref{eq:newfit} we have shown that it reduces the scatter of the
scaling relation. The scatter around the predicted relation
(Eq.~\ref{eq:newfit}) is 0.39 dex in $\log M_*^{\mathrm{DLA}}$, which
is smaller than the intrinsic scatter if we assume a constant
metallicity throughout the galaxy. $\Gamma$ has the units of
dex~kpc$^{-1}$ and in this connection represents a metallicity
gradient in the outskirts of the DLA galaxies. Some additional scatter
of metallicity gradients is expected because DLA galaxies have mixed
morphological types \citep{rao03,chen05,rao11} and random orientations
and inclinations with respect to the DLA line of sight. The
environment of the DLA galaxy may also play a role, as mergers and
interactions between galaxies will change the gradients.

\subsection{Group environments}
\label{sect:group_env}
Some low-redshift DLA galaxies are located in crowded regions, where,
on average, five separate galaxies are detected within 10\arcsec\ from
the QSO line of sight \citep{chen03}. An origin in a group environment
is consistent with the fact that some of the low-redshift DLA galaxies
discussed in this paper are found in spectroscopically confirmed
groups. This is the case for Q1127--147 and Q0235+164
\citep{rao03,kacprzak10}, while another example of a DLA galaxy known
to reside in a group is the DLA towards PKS~0952+179
\citep{rao03}. Even though spectroscopic confirmations are missing,
several other DLA galaxies reside in over-densities within 50 kpc at
$z<1$ \citep{chen03,rao11}.

For the high-redshift DLA sample, we know that the region around the
DLA galaxy towards Q2206-1958 has a complex morphology with several
compact clumps at the DLA redshift \citep{moller02,weatherley05}. As
such, it presents a clear case of a high redshift merger event and the
build up of a massive galaxy. In other cases, multiple
galaxies are detected within $\sim$6\arcsec (corresponding to 50 kpc)
from the QSO line of sight in high spatial resolution \emph{HST}
images \citep{warren01}. However, without spectroscopic observations
of all these galaxies, we cannot determine if these also represent
galaxies in groups environments at the DLA redshift, lie at the QSO
redshifts, or are entirely unrelated to either of these.

DLA sight lines that are found to intersect group environments pose
the specific question of how to relate the confirmed DLA galaxy(ies)
and the DLA gas metallicity. There is no simple and unique answer to
this question since some groups are only loosely bound and may have
been formed very recently, while other groups are very compact and
show clear signs of interaction and even merging in progress.
Interactions between galaxies give rise to more shallow metallicity
gradients \citep{rupke10}, but observations only refer to strongly
interacting spiral galaxies, while here we study the extended
outskirts of the galaxies irrespectively on their morphologies.  We
have therefore taken the view that in open groups where galaxies at
the DLA redshift are found at very large projected distances from each
other, like in the case of Q1127--145, the DLA gas metallicity is
likely to be closely connected to the galaxy at the smallest impact
parameter. In compact groups on the other hand, the DLA gas
metallicity could reveal properties of the tidally stripped gas in the
group, and therefore reflect the total integrated mass of the merging
group like in the case of Q2206--1958.

The question of possible mis-identification is addressed in
  Appendix A, where we demonstrate that the results in
  Sect.~\ref{sect:results} are robust. Our tests support that the
  strategy to identify the DLA galaxy described above is correct.  

\subsection{Metallicity gradients in absorption selected galaxies} 
While we infer metallicity gradients in DLA galaxies indirectly via
the stellar mass prediction, metallicity gradients can be measured
directly by comparing the DLA absorption metallicities with oxygen
abundances derived from strong emission-line diagnostics as discussed
in Sect.~\ref{sect:cfunct}.  We showed that the two methods both
reveal gradients around --0.02 dex kpc$^{-1}$. A few other DLAs have
measured metallicity gradients consistent with this number.
\citet{chen05} analyse six DLA systems at $z<0.65$ (four of which are
also included in our investigation, while the other two do not
  fulfill the criteria listed in Section~\ref{sect:sample}) and find
an average gradient of ($-0.041\pm0.012)h$ dex~kpc$^{-1}$. With
$h=0.7$ in the cosmology adopted here, this implies a gradient of
$-0.029\pm0.08$ dex~kpc$^{-1}$. For two DLA systems at $z\sim1$,
\citet{peroux14} find similar shallow gradients of $-0.03\pm0.08$ and
$-0.02\pm0.08$ dex~kpc$^{-1}$.

\subsection{Metallicity gradients in galaxy discs} 
 While the measurements of the gradients above refer to the extended
 outskirts of the galaxies, in few cases, metallicity gradients can be
 measured directly in the galaxy discs. This requires high
 spatial-resolution and resolved spectroscopy for example from
 integral field spectra coupled with adaptive optics
 observations. \citet{peroux11a,peroux13} use VLT/SINFONI to measure
 metallicity gradients based on spatially resolved emission line
 ratios from \ion{H}{ii} regions in the discs of three DLA galaxies
 including Q1009--0026, and they report values ranging from $-0.07$ to
 $-0.11$ dex~kpc$^{-1}$ measured over a few kpc in radius.

Investigations of non-absorption selected galaxies chosen within the
same redshift bins and luminosity intervals, have substantial larger
sample sizes to draw from. Shallow gradients of $-0.027\pm0.005$ dex
kpc$^{-1}$ in galaxy discs at $0.8<z<2.2$ have been measured
\citep{swinbank12}, while even positive gradients have been inferred
from galaxies at $z\sim1.2$ \citep{queyrel12}.

Metallicity gradients derived from emission in \ion{H}{ii} regions
within a few kpc in the discs represent the chemical evolution taking
place within the substructure of the galaxy discs themselves. Young
starbursts may quickly enrich the immediate surroundings of the
\ion{H}{ii} regions, and thereby on short time scales give rise to
steep gradients. When combining DLA metallicities with the global
galaxy metallicity integrated over scales of tens of kpc on the other
hand, we find much more shallow metallicity gradients in the outskirts
of the DLA galaxies (--0.022 dex kpc$^{-1}$). These shallow gradients
reflect more closely the long-duration time scale of galaxy evolution
and the gradual build of metals in the outer regions of galaxies, and
is affected by both infalling pristine gas and outflows of metal rich
gas.

\subsection{Absorption line gradients in galaxy halos} 
 The extended regions at distances of up to 40 kpc studied here
  probe the region where the interstellar medium extends to the
  circumgalactic medium (CGM) of galaxies. The metal composition of
  the CGM has been investigated in other types of galaxies besides DLA
  systems.  Using stacks of spectra of Lyman break galaxies
  \citet{steidel10} demonstrate that metal lines are detected out to
  distances of $\sim$100 kpc, and that the absorption equivalent
  widths (EWs) are anti-correlated with distance. In low-redshift
  ($z\sim0.2$) galaxies a similar anti-correlation is seen
  \citep{werk13}. In the CGM of Lyman-break galaxies,
  \citet{steidel10} measure a change of the rest-frame EW(\ion{Si}{ii}
  $\lambda$1526) from $\sim$2~{\AA} at 2~kpc to 0.4~{\AA} at 30
  kpc. For DLAs, there is a tight correlation between EW(\ion{Si}{ii}
  $\lambda$1526) and metallicity \citep{prochaska08}. If a similar
  relation exist for the halos around Lyman break galaxies, this
  implies an average gradient of $-0.026\pm0.012$ dex
  kpc$^{-1}$. Unfortunately, the EW of \ion{Si}{ii} is rarely reported
  for DLAs, so we cannot test this relation directly in our sample.

  \ion{Mg}{ii} absorbers are also frequently used to probe the CGM,
  showing a clear trend of decreasing EW with impact parameter
  \citep{churchill00}. However, the strongest systems with rest-frame
  EW(\ion{Mg}{ii} $\lambda$2796)$>$1{\AA} do not show a strong
  anti-correlation \citep{nielsen13}. The DLA systems analysed here
  are likely all strong \ion{Mg}{ii} absorbers \citep[see
    also][]{rao00}. For six of the DLAs, which have reported
  rest-frame EW(\ion{Mg}{ii}) between $0.6-2.7$ {\AA} in the
  literature, we do not find any clear trend with impact parameter as
  expected.


\section{Conclusions}
\label{sect:conclusions}

The long standing quest for the nature of DLA galaxies is finally
nearing its conclusion. It was recently shown (M13) that the stellar
mass of DLA galaxies can be computed from a simple relation depending
only on the metallicity of the absorbing gas, the redshift, and a
parameter, $C_{\mathrm{[M/H]}}$, which is the offset between metallicity
measured from the absorbing DLA gas and that measured from emission
lines of the same galaxy.  The $C_{\mathrm{[M/H]}}$ parameter is in
most cases not known, and in this paper we have addressed how it may be
determined or estimated in cases where it cannot be measured directly.

$C_{\mathrm{[M/H]}}$ could be a function of several properties of the
DLA galaxy, but in particular one of those is an obvious candidate. From
studies at low redshifts it is known that galaxies have metallicity
gradients such that they in general have a higher metallicity in the
centre and in the mean correspondingly
lower metallicity at increasing distance from the centre.

In this paper we have addressed two questions. Since the prescription
for computing DLA stellar masses provided in M13 was derived on a
purely statistical basis we first tested, via comparison to stellar
masses determined directly from SED fits to photometric data, if the
prescription is correct. As a byproduct of this test we also
determined individual values of  $C_{\mathrm{[M/H]}}$ for each DLA
galaxy in our sample. Secondly we then tested if our sample showed any
evidence for the expected signature of metallicity gradients. Our
conclusions on those two tests can be summarised as follows:

\begin{enumerate}
\item Our independent test confirms the statistical MzZ relation
  reported in M13. We find a mean value $C_{\mathrm{[M/H]}} =
  0.44\pm0.10$ with a scatter of $0.31$. In case nothing else than
  metallicity and redshift is known about a DLA galaxy then we
  recommend to use this value in combination with the M13
  prescription.

\item We also find that the data show a correlation between
  $C_{\mathrm{[M/H]}}$ and impact parameter similar to known
  metallicity gradients at low redshift. We find a best fit for a
  gradient $-0.022\pm 0.004$ dex~kpc$^{-1}$ in the entire range of
  redshifts $z=0.1$ to $3.2$.  The sample is still very small, and
  because of the distribution of impact parameters and redshifts in
  our sample, one could also interpret the correlation as a redshift
  evolution of $C_{\mathrm{[M/H]}}$ without metallicity gradients.
  The redshift evolution interpretation is in conflict with results
  from several large independent surveys, while the metallicity
  gradient interpretation is favoured because metallicity gradients
  are well documented in the local universe and out to at least $z=1$.

\item The residual internal scatter of the relation is significantly
reduced (for the same number of fitted parameters) in the metallicity
gradient formulation of the prescription. This is additional independent
support that we are indeed measuring metallicity gradients in the DLA
galaxies.

\item The sample includes three sub-DLA systems. If we
exclude those from the analysis, the results remain unchanged. This
suggests that high-column density sub-DLAs follow the same relation as
laid out by classical DLAs.
\end{enumerate}

Based on those results we have expanded the $C_{\mathrm{[M/H]}}$ parameter
and presented an improved prescription as given in
Eq.~\ref{eq:newfit}. We recommend to use this updated form of the
relation in cases where the impact parameter is known, or where at least
limits can be placed on it.

\section*{Acknowledgements}
We thank Roser Pello for providing updates of- and help with the use of the
{\sc HyperZ} code.  The Dark Cosmology Centre is funded by the DNRF.  LC is
supported by the EU under a Marie Curie Intra-European Fellowship, contract
PIEF-GA-2010-274117.  The research leading to these results has received
funding from the European Research Council under the European Union's Seventh
Framework Program (FP7/2007-2013)/ERC Grant agreement no. EGGS-278202.  LC
thanks ESO, Garching for the hospitality during the work for this paper, and
for the support by the DFG cluster of excellence `Origin and Structure of the
Universe'. 


\bibliographystyle{apj}
\bibliography{ms_LC}

\begin{thebibliography}{}

\bibitem[\protect\citeauthoryear{{Asplund} et~al.}{{Asplund}
  et~al.}{2009}]{asplund09}
{Asplund}, M., {Grevesse}, N., {Sauval}, A.~J.,  \& {Scott}, P. 2009, \araa,
  47, 481

\bibitem[\protect\citeauthoryear{{Belli} et~al.}{{Belli}
  et~al.}{2013}]{belli13}
{Belli}, S., {Jones}, T., {Ellis}, R.~S.,  \& {Richard}, J. 2013, \apj, 772,
  141

\bibitem[\protect\citeauthoryear{{Bergeron} \& {Boiss{\'e}}}{{Bergeron} \&
  {Boiss{\'e}}}{1991}]{bergeron91}
{Bergeron}, J.,  \& {Boiss{\'e}}, P. 1991, \aap, 243, 344

\bibitem[\protect\citeauthoryear{{Bolzonella}, {Miralles}, \&
  {Pell{\'o}}}{{Bolzonella} et~al.}{2000}]{bolzonella00}
{Bolzonella}, M., {Miralles}, J.-M.,  \& {Pell{\'o}}, R. 2000, \aap, 363, 476

\bibitem[\protect\citeauthoryear{{Bouch{\'e}} et~al.}{{Bouch{\'e}}
  et~al.}{2013}]{bouche13}
{Bouch{\'e}}, N., {Murphy}, M.~T., {Kacprzak}, G.~G., {P{\'e}roux}, C.,
  {Contini}, T., {Martin}, C.~L.,  \& {Dessauges-Zavadsky}, M. 2013, Science,
  341, 50

\bibitem[\protect\citeauthoryear{{Bouch{\'e}} et~al.}{{Bouch{\'e}}
  et~al.}{2012}]{bouche12}
{Bouch{\'e}}, N., et~al. 2012, \mnras, 419, 2

\bibitem[\protect\citeauthoryear{{Bruzual} \& {Charlot}}{{Bruzual} \&
  {Charlot}}{2003}]{bruzual03}
{Bruzual}, G.,  \& {Charlot}, S. 2003, \mnras, 344, 1000

\bibitem[\protect\citeauthoryear{{Centuri{\'o}n} et~al.}{{Centuri{\'o}n}
  et~al.}{2003}]{centurion03}
{Centuri{\'o}n}, M., {Molaro}, P., {Vladilo}, G., {P{\'e}roux}, C.,
  {Levshakov}, S.~A.,  \& {D'Odorico}, V. 2003, \aap, 403, 55

\bibitem[\protect\citeauthoryear{{Chabrier}}{{Chabrier}}{2003}]{chabrier03}
{Chabrier}, G. 2003, \pasp, 115, 763

\bibitem[\protect\citeauthoryear{{Chen}, {Kennicutt}, \& {Rauch}}{{Chen}
  et~al.}{2005}]{chen05}
{Chen}, H.-W., {Kennicutt}, R.~C., Jr.,  \& {Rauch}, M. 2005, \apj, 620, 703

\bibitem[\protect\citeauthoryear{{Chen} \& {Lanzetta}}{{Chen} \&
  {Lanzetta}}{2003}]{chen03}
{Chen}, H.-W.,  \& {Lanzetta}, K.~M. 2003, \apj, 597, 706

\bibitem[\protect\citeauthoryear{{Christensen} et~al.}{{Christensen}
  et~al.}{2012}]{christensen12}
{Christensen}, L., et~al. 2012, \mnras, 427, 1973

\bibitem[\protect\citeauthoryear{{Christensen} et~al.}{{Christensen}
  et~al.}{2004}]{christensen04}
{Christensen}, L., {S{\'a}nchez}, S.~F., {Jahnke}, K., {Becker}, T.,
  {Wisotzki}, L., {Kelz}, A., {Popovi{\'c}}, L.~{\v C}.,  \& {Roth}, M.~M.
  2004, \aap, 417, 487

\bibitem[\protect\citeauthoryear{{Christensen} et~al.}{{Christensen}
  et~al.}{2007}]{christensen07}
{Christensen}, L., {Wisotzki}, L., {Roth}, M.~M., {S{\'a}nchez}, S.~F., {Kelz},
  A.,  \& {Jahnke}, K. 2007, \aap, 468, 587

\bibitem[\protect\citeauthoryear{{Chun} et~al.}{{Chun} et~al.}{2010}]{chun10}
{Chun}, M.~R., {Kulkarni}, V.~P., {Gharanfoli}, S.,  \& {Takamiya}, M. 2010,
  \aj, 139, 296

\bibitem[\protect\citeauthoryear{{Churchill} et~al.}{{Churchill}
  et~al.}{2000}]{churchill00}
{Churchill}, C.~W., {Mellon}, R.~R., {Charlton}, J.~C., {Jannuzi}, B.~T.,
  {Kirhakos}, S., {Steidel}, C.~C.,  \& {Schneider}, D.~P. 2000, \apjs, 130, 91

\bibitem[\protect\citeauthoryear{{Cooke} et~al.}{{Cooke}
  et~al.}{2011}]{cooke11}
{Cooke}, R., {Pettini}, M., {Steidel}, C.~C., {Rudie}, G.~C.,  \& {Nissen},
  P.~E. 2011, \mnras, 417, 1534

\bibitem[\protect\citeauthoryear{{Djorgovski} et~al.}{{Djorgovski}
  et~al.}{1996}]{djorgovski96}
{Djorgovski}, S.~G., {Pahre}, M.~A., {Bechtold}, J.,  \& {Elston}, R. 1996,
  \nat, 382, 234

\bibitem[\protect\citeauthoryear{{Erb} et~al.}{{Erb} et~al.}{2006}]{erb06}
{Erb}, D.~K., {Shapley}, A.~E., {Pettini}, M., {Steidel}, C.~C., {Reddy},
  N.~A.,  \& {Adelberger}, K.~L. 2006, \apj, 644, 813

\bibitem[\protect\citeauthoryear{{Fukugita} \& {M{\'e}nard}}{{Fukugita} \&
  {M{\'e}nard}}{2014}]{fukugita14}
{Fukugita}, M.,  \& {M{\'e}nard}, B. 2014, ArXiv e-prints

\bibitem[\protect\citeauthoryear{{Fynbo} et~al.}{{Fynbo}
  et~al.}{2013}]{fynbo13}
{Fynbo}, J.~P.~U., et~al. 2013, \mnras, 436, 361

\bibitem[\protect\citeauthoryear{{Fynbo} et~al.}{{Fynbo}
  et~al.}{2010}]{fynbo10}
{Fynbo}, J.~P.~U., et~al. 2010, \mnras, 408, 2128

\bibitem[\protect\citeauthoryear{{Fynbo} et~al.}{{Fynbo}
  et~al.}{2011}]{fynbo11}
{Fynbo}, J.~P.~U., et~al. 2011, \mnras, 413, 2481

\bibitem[\protect\citeauthoryear{{Fynbo} et~al.}{{Fynbo}
  et~al.}{2008}]{fynbo08}
{Fynbo}, J.~P.~U., {Prochaska}, J.~X., {Sommer-Larsen}, J.,
  {Dessauges-Zavadsky}, M.,  \& {M{\o}ller}, P. 2008, \apj, 683, 321

\bibitem[\protect\citeauthoryear{{Guillemin} \& {Bergeron}}{{Guillemin} \&
  {Bergeron}}{1997}]{guillemin97}
{Guillemin}, P.,  \& {Bergeron}, J. 1997, \aap, 328, 499

\bibitem[\protect\citeauthoryear{{Ilbert} et~al.}{{Ilbert}
  et~al.}{2009}]{ilbert09}
{Ilbert}, O., et~al. 2009, \apj, 690, 1236

\bibitem[\protect\citeauthoryear{{Jorgenson} \& {Wolfe}}{{Jorgenson} \&
  {Wolfe}}{2014}]{jorgenson14}
{Jorgenson}, R.~A.,  \& {Wolfe}, A.~M. 2014, \apj, 785, 16

\bibitem[\protect\citeauthoryear{{Kacprzak}, {Murphy}, \&
  {Churchill}}{{Kacprzak} et~al.}{2010}]{kacprzak10}
{Kacprzak}, G.~G., {Murphy}, M.~T.,  \& {Churchill}, C.~W. 2010, \mnras, 406,
  445

\bibitem[\protect\citeauthoryear{{Kewley} \& {Dopita}}{{Kewley} \&
  {Dopita}}{2002}]{kewley02}
{Kewley}, L.~J.,  \& {Dopita}, M.~A. 2002, \apjs, 142, 35

\bibitem[\protect\citeauthoryear{{Kewley} \& {Ellison}}{{Kewley} \&
  {Ellison}}{2008}]{kewley08}
{Kewley}, L.~J.,  \& {Ellison}, S.~L. 2008, \apj, 681, 1183

\bibitem[\protect\citeauthoryear{{Komatsu} et~al.}{{Komatsu}
  et~al.}{2011}]{komatsu11}
{Komatsu}, E., et~al. 2011, \apjs, 192, 18

\bibitem[\protect\citeauthoryear{{Krogager} et~al.}{{Krogager}
  et~al.}{2013}]{krogager13}
{Krogager}, J.-K., et~al. 2013, \mnras, 433, 3091

\bibitem[\protect\citeauthoryear{{Krogager} et~al.}{{Krogager}
  et~al.}{2012}]{krogager12}
{Krogager}, J.-K., {Fynbo}, J.~P.~U., {M{\o}ller}, P., {Ledoux}, C.,
  {Noterdaeme}, P., {Christensen}, L., {Milvang-Jensen}, B.,  \& {Sparre}, M.
  2012, \mnras, 424, L1

\bibitem[\protect\citeauthoryear{{Kulkarni} et~al.}{{Kulkarni}
  et~al.}{2005}]{kulkarni05}
{Kulkarni}, V.~P., {Fall}, S.~M., {Lauroesch}, J.~T., {York}, D.~G., {Welty},
  D.~E., {Khare}, P.,  \& {Truran}, J.~W. 2005, \apj, 618, 68

\bibitem[\protect\citeauthoryear{{Kulkarni} et~al.}{{Kulkarni}
  et~al.}{2006}]{kulkarni06}
{Kulkarni}, V.~P., {Woodgate}, B.~E., {York}, D.~G., {Thatte}, D.~G.,
  {Meiring}, J., {Palunas}, P.,  \& {Wassell}, E. 2006, \apj, 636, 30

\bibitem[\protect\citeauthoryear{{Lane} et~al.}{{Lane} et~al.}{1998}]{lane98}
{Lane}, W., {Smette}, A., {Briggs}, F., {Rao}, S., {Turnshek}, D.,  \&
  {Meylan}, G. 1998, \aj, 116, 26

\bibitem[\protect\citeauthoryear{{Lara-L{\'o}pez} et~al.}{{Lara-L{\'o}pez}
  et~al.}{2010}]{lara-lopez10}
{Lara-L{\'o}pez}, M.~A., et~al. 2010, \aap, 521, L53

\bibitem[\protect\citeauthoryear{{Le Brun} et~al.}{{Le Brun}
  et~al.}{1997}]{lebrun97}
{Le Brun}, V., {Bergeron}, J., {Boisse}, P.,  \& {Deharveng}, J.~M. 1997, \aap,
  321, 733

\bibitem[\protect\citeauthoryear{{Ledoux}, {Bergeron}, \& {Petitjean}}{{Ledoux}
  et~al.}{2002}]{ledoux02}
{Ledoux}, C., {Bergeron}, J.,  \& {Petitjean}, P. 2002, \aap, 385, 802

\bibitem[\protect\citeauthoryear{{Ledoux} et~al.}{{Ledoux}
  et~al.}{2006}]{ledoux06}
{Ledoux}, C., {Petitjean}, P., {Fynbo}, J.~P.~U., {M{\o}ller}, P.,  \&
  {Srianand}, R. 2006, \aap, 457, 71

\bibitem[\protect\citeauthoryear{{Madau}}{{Madau}}{1995}]{madau95}
{Madau}, P. 1995, \apj, 441, 18

\bibitem[\protect\citeauthoryear{{Maiolino} et~al.}{{Maiolino}
  et~al.}{2008}]{maiolino08}
{Maiolino}, R., et~al. 2008, \aap, 488, 463

\bibitem[\protect\citeauthoryear{{Mannucci} et~al.}{{Mannucci}
  et~al.}{2010}]{mannucci10}
{Mannucci}, F., {Cresci}, G., {Maiolino}, R., {Marconi}, A.,  \& {Gnerucci}, A.
  2010, \mnras, 408, 2115

\bibitem[\protect\citeauthoryear{{M{\o}ller}, {Fynbo}, \& {Fall}}{{M{\o}ller}
  et~al.}{2004}]{moller04}
{M{\o}ller}, P., {Fynbo}, J.~P.~U.,  \& {Fall}, S.~M. 2004, \aap, 422, L33

\bibitem[\protect\citeauthoryear{{M{\o}ller} et~al.}{{M{\o}ller}
  et~al.}{2013}]{moller13}
{M{\o}ller}, P., {Fynbo}, J.~P.~U., {Ledoux}, C.,  \& {Nilsson},
K.~K. 2013 (M13),
  \mnras, 430, 2680

\bibitem[\protect\citeauthoryear{{M{\o}ller} \& {Jakobsen}}{{M{\o}ller} \&
  {Jakobsen}}{1990}]{moller90}
{M{\o}ller}, P.,  \& {Jakobsen}, P. 1990, \aap, 228, 299

\bibitem[\protect\citeauthoryear{{M{\o}ller} \& {Warren}}{{M{\o}ller} \&
  {Warren}}{1998}]{moller98}
{M{\o}ller}, P.,  \& {Warren}, S.~J. 1998, \mnras, 299, 661

\bibitem[\protect\citeauthoryear{{M{\o}ller} et~al.}{{M{\o}ller}
  et~al.}{2002}]{moller02}
{M{\o}ller}, P., {Warren}, S.~J., {Fall}, S.~M., {Fynbo}, J.~U.,  \&
  {Jakobsen}, P. 2002, \apj, 574, 51

\bibitem[\protect\citeauthoryear{{Neeleman} et~al.}{{Neeleman}
  et~al.}{2013}]{neeleman13}
{Neeleman}, M., {Wolfe}, A.~M., {Prochaska}, J.~X.,  \& {Rafelski}, M. 2013,
  \apj, 769, 54

\bibitem[\protect\citeauthoryear{{Nielsen}, {Churchill}, \&
  {Kacprzak}}{{Nielsen} et~al.}{2013}]{nielsen13}
{Nielsen}, N.~M., {Churchill}, C.~W.,  \& {Kacprzak}, G.~G. 2013, \apj, 776,
  115

\bibitem[\protect\citeauthoryear{{Noeske} et~al.}{{Noeske}
  et~al.}{2007}]{noeske07}
{Noeske}, K.~G., et~al. 2007, \apjl, 660, L43

\bibitem[\protect\citeauthoryear{{Noterdaeme} et~al.}{{Noterdaeme}
  et~al.}{2012}]{noterdaeme12}
{Noterdaeme}, P., et~al. 2012, \aap, 540, A63

\bibitem[\protect\citeauthoryear{{P{\'e}roux} et~al.}{{P{\'e}roux}
  et~al.}{2013}]{peroux13}
{P{\'e}roux}, C., {Bouch{\'e}}, N., {Kulkarni}, V.~P.,  \& {York}, D.~G. 2013,
  \mnras, 436, 2650

\bibitem[\protect\citeauthoryear{{P{\'e}roux} et~al.}{{P{\'e}roux}
  et~al.}{2011a}]{peroux11a}
{P{\'e}roux}, C., {Bouch{\'e}}, N., {Kulkarni}, V.~P., {York}, D.~G.,  \&
  {Vladilo}, G. 2011a, \mnras, 410, 2237

\bibitem[\protect\citeauthoryear{{P{\'e}roux} et~al.}{{P{\'e}roux}
  et~al.}{2011b}]{peroux11b}
{P{\'e}roux}, C., {Bouch{\'e}}, N., {Kulkarni}, V.~P., {York}, D.~G.,  \&
  {Vladilo}, G. 2011b, \mnras, 410, 2251

\bibitem[\protect\citeauthoryear{{P{\'e}roux} et~al.}{{P{\'e}roux}
  et~al.}{2012}]{peroux12}
{P{\'e}roux}, C., {Bouch{\'e}}, N., {Kulkarni}, V.~P., {York}, D.~G.,  \&
  {Vladilo}, G. 2012, \mnras, 419, 3060

\bibitem[\protect\citeauthoryear{{P{\'e}roux}, {Kulkarni}, \&
  {York}}{{P{\'e}roux} et~al.}{2014}]{peroux14}
{P{\'e}roux}, C., {Kulkarni}, V.~P.,  \& {York}, D.~G. 2014, \mnras, 437, 3144

\bibitem[\protect\citeauthoryear{{Pettini} et~al.}{{Pettini}
  et~al.}{1994}]{pettini94}
{Pettini}, M., {Smith}, L.~J., {Hunstead}, R.~W.,  \& {King}, D.~L. 1994, \apj,
  426, 79

\bibitem[\protect\citeauthoryear{{Pontzen} et~al.}{{Pontzen}
  et~al.}{2008}]{pontzen08}
{Pontzen}, A., et~al. 2008, \mnras, 390, 1349

\bibitem[\protect\citeauthoryear{{Prochaska} et~al.}{{Prochaska}
  et~al.}{2008}]{prochaska08}
{Prochaska}, J.~X., {Chen}, H.-W., {Wolfe}, A.~M., {Dessauges-Zavadsky}, M.,
  \& {Bloom}, J.~S. 2008, \apj, 672, 59

\bibitem[\protect\citeauthoryear{{Prochaska} et~al.}{{Prochaska}
  et~al.}{2003}]{prochaska03}
{Prochaska}, J.~X., {Gawiser}, E., {Wolfe}, A.~M., {Castro}, S.,  \&
  {Djorgovski}, S.~G. 2003, \apjl, 595, L9

\bibitem[\protect\citeauthoryear{{Queyrel} et~al.}{{Queyrel}
  et~al.}{2012}]{queyrel12}
{Queyrel}, J., et~al. 2012, \aap, 539, A93

\bibitem[\protect\citeauthoryear{{Rafelski} et~al.}{{Rafelski}
  et~al.}{2012}]{rafelski12}
{Rafelski}, M., {Wolfe}, A.~M., {Prochaska}, J.~X., {Neeleman}, M.,  \&
  {Mendez}, A.~J. 2012, \apj, 755, 89

\bibitem[\protect\citeauthoryear{{Rao} et~al.}{{Rao} et~al.}{2011}]{rao11}
{Rao}, S.~M., {Belfort-Mihalyi}, M., {Turnshek}, D.~A., {Monier}, E.~M.,
  {Nestor}, D.~B.,  \& {Quider}, A. 2011, \mnras, 416, 1215

\bibitem[\protect\citeauthoryear{{Rao} et~al.}{{Rao} et~al.}{2003}]{rao03}
{Rao}, S.~M., {Nestor}, D.~B., {Turnshek}, D.~A., {Lane}, W.~M., {Monier},
  E.~M.,  \& {Bergeron}, J. 2003, \apj, 595, 94

\bibitem[\protect\citeauthoryear{{Rao} \& {Turnshek}}{{Rao} \&
  {Turnshek}}{2000}]{rao00}
{Rao}, S.~M.,  \& {Turnshek}, D.~A. 2000, \apjs, 130, 1

\bibitem[\protect\citeauthoryear{{Rupke}, {Kewley}, \& {Chien}}{{Rupke}
  et~al.}{2010}]{rupke10}
{Rupke}, D.~S.~N., {Kewley}, L.~J.,  \& {Chien}, L.-H. 2010, \apj, 723, 1255

\bibitem[\protect\citeauthoryear{{Savaglio} et~al.}{{Savaglio}
  et~al.}{2005}]{savaglio05}
{Savaglio}, S., et~al. 2005, \apj, 635, 260

\bibitem[\protect\citeauthoryear{{Schaerer} \& {de Barros}}{{Schaerer} \& {de
  Barros}}{2009}]{schaerer09}
{Schaerer}, D.,  \& {de Barros}, S. 2009, \aap, 502, 423

\bibitem[\protect\citeauthoryear{{Schlafly} \& {Finkbeiner}}{{Schlafly} \&
  {Finkbeiner}}{2011}]{schlafly11}
{Schlafly}, E.~F.,  \& {Finkbeiner}, D.~P. 2011, \apj, 737, 103

\bibitem[\protect\citeauthoryear{{Schulte-Ladbeck} et~al.}{{Schulte-Ladbeck}
  et~al.}{2005}]{schulte-ladbeck05}
{Schulte-Ladbeck}, R.~E., {K{\"o}nig}, B., {Miller}, C.~J., {Hopkins}, A.~M.,
  {Drozdovsky}, I.~O., {Turnshek}, D.~A.,  \& {Hopp}, U. 2005, \apjl, 625, L79

\bibitem[\protect\citeauthoryear{{Steidel} et~al.}{{Steidel}
  et~al.}{2010}]{steidel10}
{Steidel}, C.~C., {Erb}, D.~K., {Shapley}, A.~E., {Pettini}, M., {Reddy}, N.,
  {Bogosavljevi{\'c}}, M., {Rudie}, G.~C.,  \& {Rakic}, O. 2010, \apj, 717, 289

\bibitem[\protect\citeauthoryear{{Steidel}, {Pettini}, \& {Hamilton}}{{Steidel}
  et~al.}{1995}]{steidel95}
{Steidel}, C.~C., {Pettini}, M.,  \& {Hamilton}, D. 1995, \aj, 110, 2519

\bibitem[\protect\citeauthoryear{{Swinbank} et~al.}{{Swinbank}
  et~al.}{2012}]{swinbank12}
{Swinbank}, A.~M., {Sobral}, D., {Smail}, I., {Geach}, J.~E., {Best}, P.~N.,
  {McCarthy}, I.~G., {Crain}, R.~A.,  \& {Theuns}, T. 2012, \mnras, 426, 935

\bibitem[\protect\citeauthoryear{{Tremonti} et~al.}{{Tremonti}
  et~al.}{2004}]{tremonti04}
{Tremonti}, C.~A., et~al. 2004, \apj, 613, 898

\bibitem[\protect\citeauthoryear{{Turnshek} et~al.}{{Turnshek}
  et~al.}{2001}]{turnshek01}
{Turnshek}, D.~A., {Rao}, S., {Nestor}, D., {Lane}, W., {Monier}, E.,
  {Bergeron}, J.,  \& {Smette}, A. 2001, \apj, 553, 288

\bibitem[\protect\citeauthoryear{{Warren} \& {M{\o}ller}}{{Warren} \&
  {M{\o}ller}}{1996}]{warren96}
{Warren}, S.~J.,  \& {M{\o}ller}, P. 1996, \aap, 311, 25

\bibitem[\protect\citeauthoryear{{Warren} et~al.}{{Warren}
  et~al.}{2001}]{warren01}
{Warren}, S.~J., {M{\o}ller}, P., {Fall}, S.~M.,  \& {Jakobsen}, P. 2001,
  \mnras, 326, 759

\bibitem[\protect\citeauthoryear{{Watson} et~al.}{{Watson}
  et~al.}{2011}]{watson11}
{Watson}, D., et~al. 2011, \apj, 741, 58

\bibitem[\protect\citeauthoryear{{Weatherley} et~al.}{{Weatherley}
  et~al.}{2005}]{weatherley05}
{Weatherley}, S.~J., {Warren}, S.~J., {M{\o}ller}, P., {Fall}, S.~M., {Fynbo},
  J.~U.,  \& {Croom}, S.~M. 2005, \mnras, 358, 985

\bibitem[\protect\citeauthoryear{{Werk} et~al.}{{Werk} et~al.}{2013}]{werk13}
{Werk}, J.~K., {Prochaska}, J.~X., {Thom}, C., {Tumlinson}, J., {Tripp}, T.~M.,
  {O'Meara}, J.~M.,  \& {Peeples}, M.~S. 2013, \apjs, 204, 17

\bibitem[\protect\citeauthoryear{{Wolfe}, {Gawiser}, \& {Prochaska}}{{Wolfe}
  et~al.}{2003}]{wolfe03}
{Wolfe}, A.~M., {Gawiser}, E.,  \& {Prochaska}, J.~X. 2003, \apj, 593, 235

\bibitem[\protect\citeauthoryear{{Wolfe} et~al.}{{Wolfe}
  et~al.}{2004}]{wolfe04}
{Wolfe}, A.~M., {Howk}, J.~C., {Gawiser}, E., {Prochaska}, J.~X.,  \& {Lopez},
  S. 2004, \apj, 615, 625

\bibitem[\protect\citeauthoryear{{Wuyts} et~al.}{{Wuyts}
  et~al.}{2012}]{wuyts12}
{Wuyts}, E., {Rigby}, J.~R., {Sharon}, K.,  \& {Gladders}, M.~D. 2012, \apj,
  755, 73

\bibitem[\protect\citeauthoryear{{Yanny}, {York}, \& {Gallagher}}{{Yanny}
  et~al.}{1989}]{yanny89}
{Yanny}, B., {York}, D.~G.,  \& {Gallagher}, J.~S. 1989, \apj, 338, 735

\bibitem[\protect\citeauthoryear{{Yuan} \& {Kewley}}{{Yuan} \&
  {Kewley}}{2009}]{yuan09}
{Yuan}, T.-T.,  \& {Kewley}, L.~J. 2009, \apjl, 699, L161

\bibitem[\protect\citeauthoryear{{Zaritsky}, {Kennicutt}, \&
  {Huchra}}{{Zaritsky} et~al.}{1994}]{zaritsky94}
{Zaritsky}, D., {Kennicutt}, R.~C., Jr.,  \& {Huchra}, J.~P. 1994, \apj, 420,
  87

\end{thebibliography}

\appendix

\section{Robustness of results considering group environments}
\label{sect:appendix}
In seven cases the fields included a second possible identification of
the DLA galaxy. In two of those cases the spectroscopic redshifts are
identical, in one case a divergent photometric redshift has been
published, in the remaining cases no redshift information is
available. Those multi-candidate fields represent either chance
alignments at vastly different redshifts, or could be various stages
of open or compact group environments and/or mergers in progress. In
section~\ref{sect:group_env} we described in detail the strategy we
followed in order to identify the most likely candidate. Here we will
test how robust our results are against possible mis-identifications.

\begin{figure*}
\begin{center}
\includegraphics[width=16.cm, bb=45 238 509 706,clip]{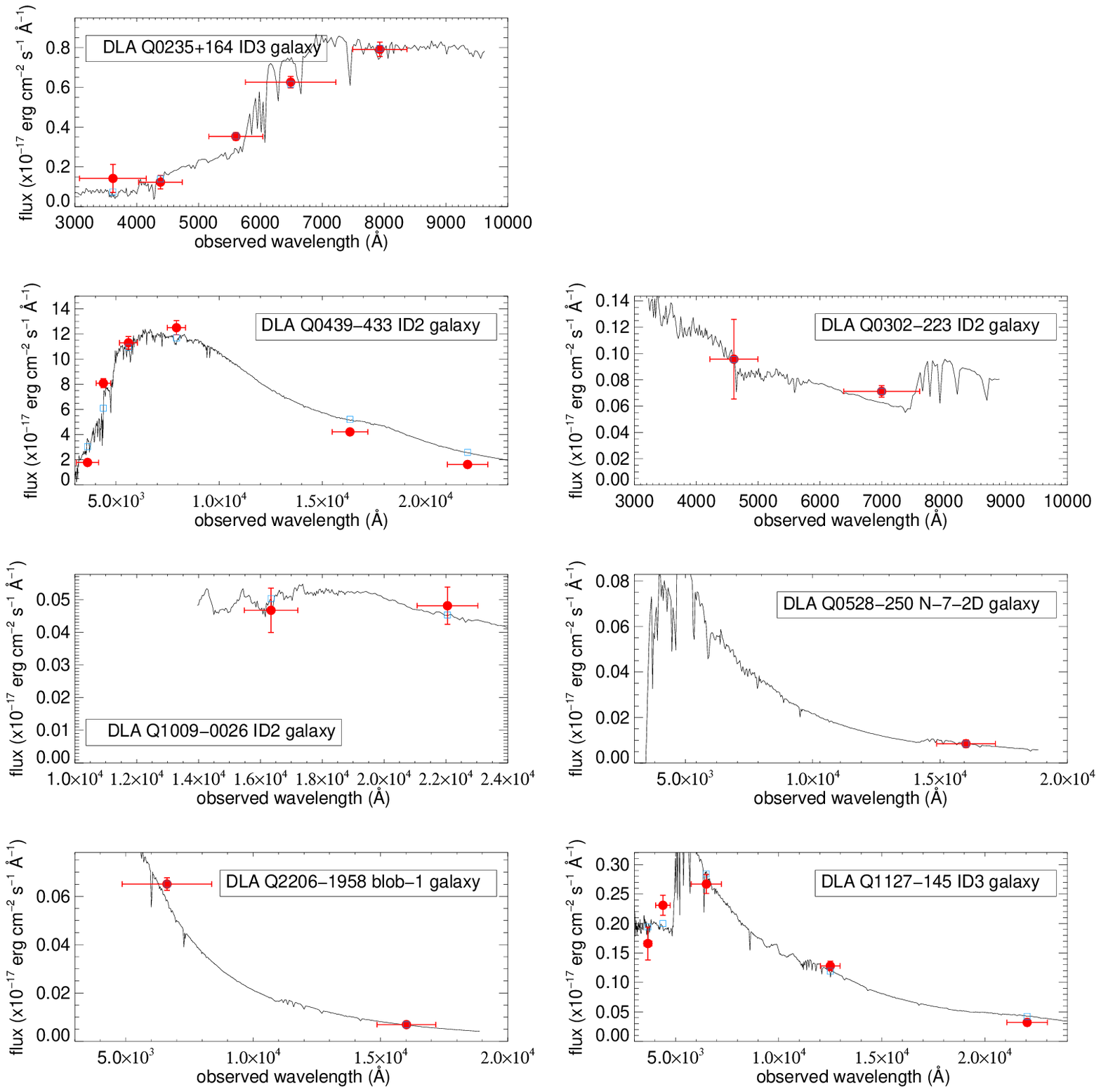}
\end{center}
\caption{Best fit SED models for the alternative candidate DLA
  galaxies assuming that they are at the redshift of the DLA.  The red
  dots denote the measured photometric points and the blue open
  squares are the expected flux density from the best fit SED model. }
\label{fig:append_fig}
\end{figure*}

\begin{figure*}
\begin{center}
\includegraphics[bb=33 342 839 1033,clip, width=18.cm]{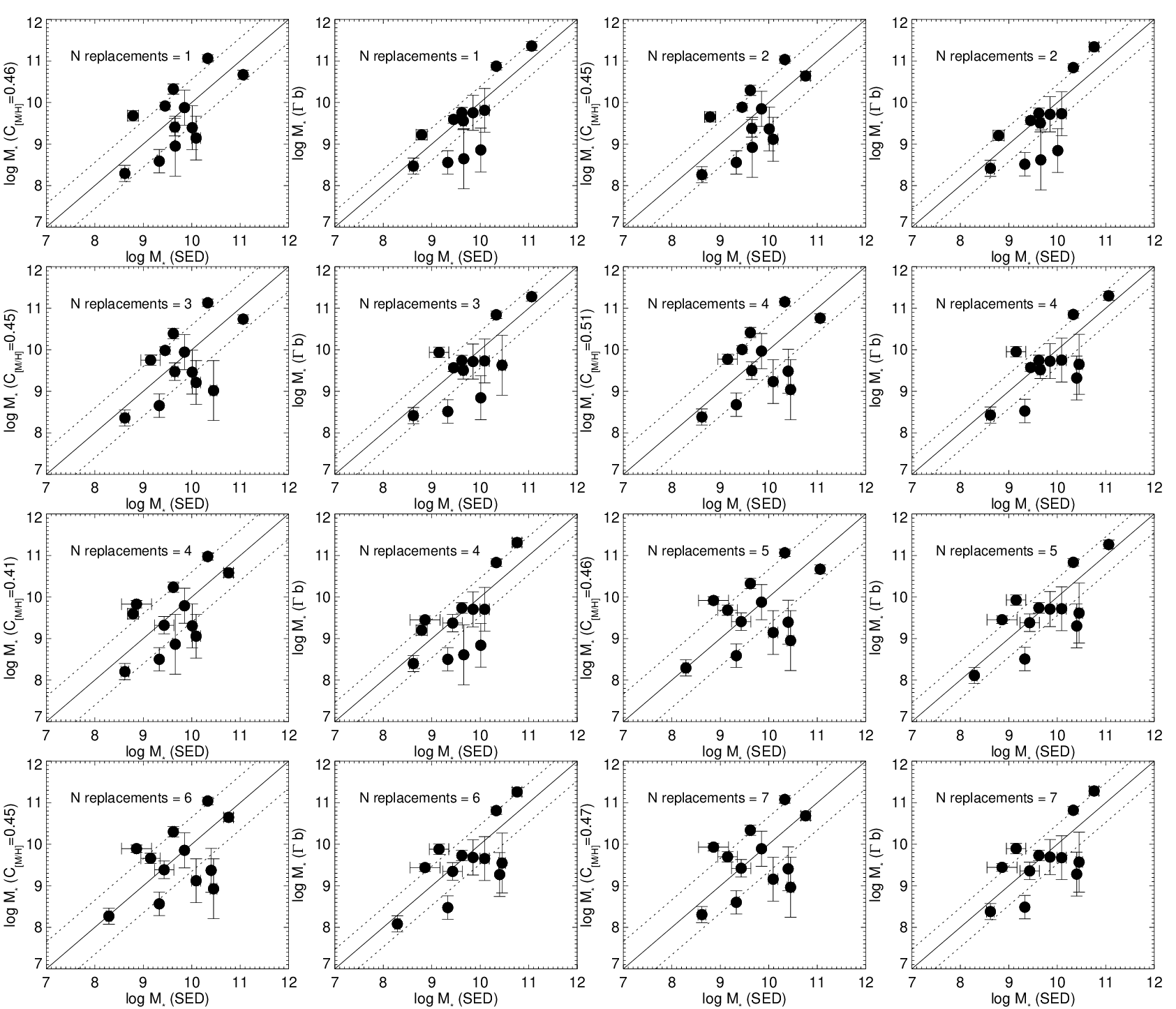}
\end{center}
\caption{Predicted stellar masses of DLA galaxies based where a random
  set of $N$ DLA galaxy candidates have been replaced. The panels in
  first and third columns are similar to Fig.~\ref{fig:mcomp} apart
  from a different offset of $C_{\rm {[M/H]}}$.  The panels in second
  and fourth columns use the corresponding best fit slope $\Gamma$b,
  and are similar to Fig 3. The scatter is increasing for increasing
  value of $N$ (Table A2).
}
\label{fig:app_mass}
\end{figure*}

In Fig.~\ref{fig:append_fig} we show the SED fits of the seven
alternative candidate DLA galaxies on the assumption that the
candidates are indeed at the redshift of the DLAs, and in
Table~\ref{tab:other_SED} we list the relevant data for those seven
galaxies. In case we assume that our strategy for identification is
incorrect then we know nothing about how to correctly identify those
seven DLAs and we have in fact just selected random galaxies. In the
mean we would therefore have 3--4 correct identifications and a
similar number of mis-identifications. At much lower probability we
might even have 5, 6, or even all 7 mis-identified.  We carried out a
full set of tests as follows. Assuming $N$ wrong identifications we
randomly picked $N$ of the candidates from Table~\ref{tab:other_SED}
and replaced the corresponding entries for $b$ and $\log M_*$ in
Tables~\ref{tab:DLAdata} and \ref{tab:gradients}.  We then determined
the mean of the results from the randomly replaced samples for all
possible values of $N$ as listed in Table~\ref{tab:app_replace}. We
found that the values of both $C_{\rm [M/H]}$ and $\Gamma$ for all $N$
are well within the 1$\sigma$ range of the results reported in
Section~\ref{sect:results}. We also found that the scatter of the
relations were monotonically rising the more of the DLA galaxies we
replaced with alternative identifications.  In Fig.~\ref{fig:app_mass}
we show a number of representative randomly replaced samples for
different values of $N$.

Based on those tests we conclude that:

1) The results reported in the main text (value of $C_{\rm [M/H]}$,
metallicity slope and general validity of Eq. 1) are very robust
against the possible mis-identifications.

2) The more mis-identifications we assume the larger is the resulting
scatter of the relations. This strongly supports that the strategy
for identification we set out in Sect.~\ref{sect:group_env} is correct.

\begin{table}
\begin{tabular}{lrrl}
\hline
\hline
name   &  $b$ [kpc]   & log $M_*$(SED) [M$_{\odot}$] & $z$ information\\ 
\hline
Q0235+164 ID3       &   40.9 &  10.45$\pm$0.07  & same $z_{\rm spec}$\\
Q0302--223 ID2      &   21.8 &   9.43$\pm$0.20  & no $z$ info\\
Q0439--433 ID2      &   20.0 &  10.40$\pm$0.01  & $z_{\rm phot}$ = 0.8\\
Q0528--250 N-7-2D   &   29.3 &   9.15$\pm$0.20  & no $z$ info \\
Q1009--0026 ID2     &   40.7 &  10.76$\pm$0.10  & no $z$ info\\
Q1127--145 ID3      &   25.7 &   8.62$\pm$0.08  & no $z$ info\\
Q2206--1958 blob2   &    9.7 &   8.86$\pm$0.31  & same $z_{\rm spec}$\\
\hline
\end{tabular}
\caption{Alternative candidate DLA galaxies in the field around the
  QSOs. SED fits were carried out as in Section~\ref{sect:SEDfits}
  fixing the redshift to that of the DLAs. The mass of Q0528--250
  N-7-2D was determined by scaling the $H$-band magnitude for the two
  galaxies in the field \citep{warren01} and assuming the same
  colours. }
\label{tab:other_SED}
\end{table}

\begin{table}
\begin{tabular}{ll | ll}
\hline
\hline
$N$ replacements & $\langle \sigma_{\rm nat} \rangle $ & 
$\langle \Gamma \rangle$ & $\langle \sigma_{\rm nat} \rangle $ \\
\hline
0 & 0.310  &  0.022 & 0.222\\     
1 & 0.319  &  0.021 & 0.229\\     
2 & 0.327  &  0.021 & 0.236\\     
3 & 0.336  &  0.021 & 0.244\\     
4 & 0.345  &  0.021 & 0.250\\
5 & 0.355  &  0.021 & 0.256\\
6 & 0.365  &  0.020 & 0.262\\
7 & 0.375  &  0.020 & 0.268\\
\hline
\end{tabular}
\caption{The first two columns give the average scatter on $C_{\rm
    [M/H]}$, derived by performing a number of replacements as
  described in the text.  The last two columns give the average slope
  and average scatter assuming $\Gamma b$ in Eq.~\ref{eq:newfit}.  }
\label{tab:app_replace}
\end{table}

\end{document}